\documentclass[a4paper,11pt]{article}
\pdfoutput=1
\usepackage[utf8]{inputenc}
\usepackage{graphicx}
\usepackage{jcappub} 
\usepackage{lineno}
\usepackage{xcolor}
\usepackage{hyperref}

\title{Interpreting the HI 21-cm cosmology maps through Largest Cluster Statistics -- III: Impact of the lightcone effect}

\author[a,b,c]{Hemanth Potluri,}
\author[a]{Manas Mohit Dosibhatla,}
\author[a]{Samit Kumar Pal,}
\author[a]{Leon Noble,}
\author[a,d]{Chandra Shekhar Murmu,}
\author[a]{Suman Majumdar,}
\author[e]{Saswata Dasgupta,}
\author[f,g]{Satadru Bag,}
\author[a]{Abhirup Datta}

\affiliation[a]{
Department of Astronomy, Astrophysics and Space Engineering,
Indian Institute of Technology Indore, Khandwa Road, Indore - 453552, India
}

\affiliation[b]{
Department of Physics, Stellenbosch University, Matieland 7602, South Africa
}

\affiliation[c]{
Kapteyn Astronomical Institute, University of Groningen, PO Box 800, NL-9700 AV Groningen, the Netherlands
}

\affiliation[d]{
Astrophysics Research Center of the Open University (ARCO) \& Department of Natural Sciences,
The Open University of Israel,
1 University Road, Ra'anana 4353701, Israel
}

\affiliation[e]{
Institute of Astronomy \& Kavli Institute for Cosmology, University of Cambridge,
Madingley Road, Cambridge CB3 0HA, United Kingdom
}

\affiliation[f]{
Physics Department, TUM School of Natural Sciences,
Technical University of Munich,
James-Franck-Straße 1, 85748 Garching, Germany
}

\affiliation[g]{
Max-Planck-Institut f{\"u}r Astrophysik,
Karl-Schwarzschild Straße 1, 85748 Garching, Germany
}       

\emailAdd{hemanthpotluri.98@gmail.com}
\emailAdd{dosibhatla.mohit@gmail.com}

\abstract{The redshifted 21-cm signal emitted by neutral Hydrogen (HI) is a promising probe to understand the evolution of the topology of ionized regions during the Epoch of Reionization (EoR). The topology of ionized regions allows us to infer the nature and properties of ionizing sources, i.e., early galaxies and AGNs. Traditional Fourier statistics, such as the power spectrum, help us quantify the strength of fluctuations in this field at different length scales but do not preserve its phase information. Analyzing the 21-cm brightness temperature field in the image domain retains its non-Gaussian characteristics and morphological information. One such approach is to track the coalescence of multiple ionized regions to form one contiguous ionized region spanning the universe. This is referred to as percolation, and its onset is quantified by a sharp rise in the value of the Largest Cluster Statistic (LCS) approaching unity. In this work, we carry out a percolation analysis of 21-cm brightness temperature maps by studying the redshift evolution of the LCS along a lightcone to distinguish between several simulated reionization scenarios. We have extended previous results on reionization model comparison from the analysis of coeval 21-cm maps to understand how the lightcone effect biases the observed percolation behavior and affects the distinguishability of the source models. We estimate the LCS of subvolumes of different sizes in mock SKA-Low 21-cm lightcone maps and study their redshift evolution for different reionization scenarios using a moving volume approach. We find that the optimal size of the moving volume that reproduces the percolation transition from coeval maps varies across reionization scenarios and the evolution of LCS remains distinct across the reionization scenarios for a fixed moving volume bandwidth. The inferred percolation transition strongly depends on the synthesized beam width and the percentage gain calibration error, and is tractable for an optimistic SKA-Low observational setup.}


\begin{document}
\maketitle
\flushbottom

\section{Introduction}
\label{sec:intro}

The Epoch of Reionization (EoR) refers to the period in the history of the universe when UV and X-ray photons emitted by the first galaxies and active galactic nuclei (AGN) ionized the neutral atomic hydrogen (HI) in the intergalactic medium (IGM) \citep{BARKANA2001125, Furlanetto_2004, 10.1111/j.1365-2966.2006.10502.x, 2010Natur.468...49R, DAYAL20181}. It was the last major phase transition of hydrogen in the universe. Studying the EoR provides key information to understand the evolution of the universe from its early stages with tiny fluctuations in the matter distribution observed through the cosmic microwave background (CMB) radiation \citep{2014A&A...571A..16P}, to the highly clustered non-Gaussian matter distribution observed in low-redshift galaxy surveys \citep{Bharadwaj_2000, 10.1111/j.1365-2966.2007.12035.x}. In addition, the reionization and the associated photo-heating of IGM gas from the ultraviolet background (UVB) affect subsequent galaxy formation~\citep{10.1093/mnras/stt474, 10.1093/mnras/stab602, 10.1093/mnras/stab877, Schmidt_2017_EoR-impact-on-galaxies}.

Observations of quasar spectra \citep{2006AJ....132..117F, 10.1093/mnras/stu2646, 2018Natur.553..473B}, Thomson scattering optical depth of CMB photons \citep{2020A&A...641A...6P}, and the luminosity and clustering properties of Lyman alpha (Ly$\alpha$) emitters \citep{Hu_2010, Morales_2021, 10.1093/mnras/stad2763, 2024NatAs...8..384W}, suggest that reionization is an extended process spanning roughly the redshift range of $6 \lesssim z \lesssim 15$. However, our knowledge about the properties of the sources responsible for the reionization of the universe and the evolution of the topology of ionized regions during this period still remains uncertain. The types of sources responsible for reionization (galaxies and AGNs) and their relative contributions need to be understood in detail to describe reionization completely. The 21-cm line emission from HI atoms resulting from the transition between the triplet and singlet hyperfine states of its ground state is a direct tracer of the HI distribution in the universe. Constructing tomographic maps of the brightness temperature fluctuations of the 21-cm line using a radio interferometer allows us to identify ionized regions at different redshifts from the cold spots in these maps \citep{1997ApJ...475..429M, Furlanetto_2004, 2013ASSL..396...45Z, mellema_2015_EoR_imaging}. This directly tracks the evolution of the topology of ionized regions \citep{10.1093/mnras/sty714, 10.1093/mnras/stz532}, which are expected to differ across reionization scenarios \citep{Pathak_2022}.

Direct detection of the 21-cm intensity mapping signal from the EoR has not been achieved yet because of observational challenges such as astrophysical foregrounds that are $4$--$5$ orders of magnitude larger than the actual signal \citep{10.1111/j.1365-2966.2004.08604.x, 10.1111/j.1365-2966.2008.13634.x, 10.1111/j.1365-2966.2010.17407.x, 10.1111/j.1365-2966.2011.18439.x, 10.1093/mnrasl/slx066, 10.1093/mnras/stz2533, 10.1093/mnras/staa1317}, coupled with residual calibration errors \citep{10.1093/mnras/stw1380, 10.1093/mnras/stw2277, 10.1093/mnras/stx1221, 10.1093/mnras/stac1994}, ionospheric phase errors \citep{10.1093/mnras/stx1797, Trott_2018, Pal_2025}, and other instrumental systematics \citep{Kern_2019}. However, upper limits of the 21-cm power spectrum have been reported through observations using several radio interferometric arrays such as the Giant Metrewave Radio Telescope (GMRT) \citep{10.1093/mnras/stt753}, Precision Array to Probe the Epoch of Reionization (PAPER) \citep{2014ApJ...788..106P, 2015ApJ...809...61A}, LOw Frequency ARray (LOFAR) \citep{2017ApJ...838...65P, 2020MNRAS.493.1662M, 10.1093/mnrasl/slae078, Mertens25}, Murchison Widefield Array (MWA) \citep{Barry_2019, 2020MNRAS.493.4711T, Nunhokee_2025}, and Hydrogen Epoch of Reionization Array (HERA) \citep{2023ApJ...945..124H}. These upper limits have been used to put constraints on the thermal and ionization history during the EoR and to rule out several exotic reionization models \citep{ghara_2020_IGM_constraints, mondal_2020_LOFAR_constraints, greig_2021_LOFAR_constraints, abdurashidova_2022_HERA_constraints}. Upper limits have also been reported for higher-order signal statistics such as the bispectrum using the MWA \citep{2025arXiv250704964S}. The low-frequency component of the upcoming Square Kilometre Array (SKA-Low), situated in Western Australia, will have enough sensitivity to construct tomographic maps of the EoR 21-cm signal \citep{2013ExA....36..235M, Koopmans:2015K0, 2017MNRAS.464.2234G}.

In order to interpret the observed 21-cm signal from the EoR, the signal statistics need to be modeled and compared with the observed statistics for reionization parameter estimation \citep{10.1093/mnras/stx3292, 10.1093/mnras/stab3215, Tiwari_2022, 10.1093/mnras/stad3849, Choudhury_2024, raghu_inference, 2025arXiv250813261M} or model selection \citep{10.1093/mnras/stz1297, 2025arXiv250208152B}. Even though the power spectrum, as a summary statistic, quantifies the contribution of different length scales to the signal fluctuations, it can only completely describe Gaussian fields. The EoR 21-cm signal is non-Gaussian \citep{10.1111/j.1365-2966.2005.08836.x, 10.1111/j.1365-2966.2006.10502.x, 10.1111/j.1365-2966.2006.10919.x}, and hence higher-order statistics such as the bispectrum need to be estimated for a more adequate description of the signal \citep{10.1093/mnras/stw482, 10.1093/mnras/sty535, 10.1093/mnras/stab2900, 10.1093/mnras/stae492, Noble_2024}.

Fourier domain statistics, however, cannot preserve the full information of the phases of the signal in Fourier space. This leads to an information loss on the topology of the ionized bubbles in the IGM, and how the ionized regions evolve, grow, and merge to complete the reionization process. This necessitates analyzing the tomographic intensity maps in the image domain to recover more information. Previous efforts at such analysis of 21-cm maps include topology quantifiers such as the Minkowski Functionals \citep{10.1111/j.1365-2966.2011.18219.x, 2014JKAS...47...49H, 10.1093/mnras/stw2701, 10.1093/mnras/sty714, 10.1093/mnras/stz532, 10.1093/mnras/stab1555}, Shapefinders \citep{Sahni_1998_shapefinders}, Minkowski Tensors \citep{Kapahtia_2019}, Betti numbers \citep{Kapahtia_2021}, granulometry \citep{10.1093/mnras/stx1568}, persistence theory \citep{10.1093/mnras/stz908}, etc. There have also been multiple efforts to characterize EoR 21-cm maps at the field level \citep{gillet_2019_FLI, Hassan_2020_FLI, Hortua_2020_FLI, prelogovic_2021_FLI, Zhao_2022_FLI, Neutsch_2022_FLI, ore_2025_FLI, Schosser_2025_FLI, posture_2025_FLI}. An effective way to track the evolution of ionized bubbles is through percolation theory \citep{1983PAZh....9..195S, 1993ApJ...413...48K, Bharadwaj_2000, Furlanetto_2016_percolation, 10.1093/mnras/sty714, 10.1093/mnras/stz532, Pathak_2022, Dasgupta_2023, Regos_2024, samit_lcs}. This is based on probing the percolation transition of the ionized bubbles, which marks the stage where the small ionized bubbles coalesce to form one large ionized region spanning a simulation box. Extending this with periodic boundary conditions, the largest ionized region (LIR) can be considered to span the entire universe, culminating in large-scale connectivity of ionized regions.

Instead of detecting a large number of ionized regions, detecting a few large ionized regions in the simulation box is sufficient to track the percolation process using the Largest Cluster Statistic (LCS) \citep{10.1093/mnras/sty714}. Previous work by Pathak et al. [\citealp[\textbf{Paper I} hereafter]{Pathak_2022}] has demonstrated that the multiple reionization source models of \citep{majumdar_2016_source-models} with the same global reionization history that can lead to different progressions of reionization (inside-out and outside-in \citep{TRC_inside-out}) can be distinguished using the LCS. Among previous papers in this series, Dasgupta et al. [\citealp[\textbf{Paper IIa} hereafter]{Dasgupta_2023}] studied how the convolution of the target signal with the synthesized beam of SKA-Low affects the inferred percolation transition point. Pal et al. [\citealp[\textbf{Paper IIb} hereafter]{samit_lcs}] studied the tolerance level to gain calibration errors in the recovery of the percolation behavior of foreground-contaminated maps.

Papers I, IIa, and IIb utilized coeval simulation boxes for estimating the signal statistics. In reality, the observed signal in different frequency channels originates from different redshifts, i.e., different cosmic times, due to the finite time light takes to travel and the expansion of the universe. As a result, the signal evolves across every slice along the line of sight of a tomographic map, introducing biases relative to coeval boxes. This is known as the lightcone effect, and its impact has been studied on the 21-cm power spectrum \citep{10.1111/j.1365-2966.2012.21293.x, 10.1093/mnras/stu927, Plante_2014, 10.1093/mnras/stu035, 10.1093/mnras/stv1855, 10.1093/mnras/stx2888, 10.1093/mnras/stab2347} and bispectrum \citep{10.1093/mnras/stab2900}. In this work, we study the bias introduced by the lightcone effect on the inferred percolation transition point for the different reionization scenarios. We aim to determine the optimal size of the chunks to be cut out of the lightcone maps to estimate the LCS. To do this, we attempt to strike a balance between a large enough chunk for a robust estimation of the LCS and a small enough chunk to minimize the redshift evolution of the signal along the line of sight. Additionally, we also study the evolution of LCS computed at a fixed chunk size to understand the distinguishability of reionization scenarios. We subsequently probe the recovery prospects of the percolation transition in SKA-Low 21-cm tomographic lightcone maps by incorporating approximate telescopic beam effects, thermal noise, and realistic gain calibration errors.

This work is structured as follows: section \ref{sec:simulations} describes the simulations of the 21-cm lightcone maps, section \ref{sec:methodology} outlines the methodology followed for the percolation analysis, and section \ref{sec:results} describes the results of the percolation analysis on lightcone maps. We summarize and conclude the work in section \ref{sec:summary}. Throughout this work, we use the the $\Lambda$CDM cosmology of the WMAP five-year data release: $h = 0.7$, $\Omega_m = 0.27$, $\Omega_\Lambda = 0.73$, $\Omega_b h^2 = 0.0226$.

\section{Simulations}
\label{sec:simulations}

\subsection{Semi-numerical simulation of reionization}
Theoretical modeling of the reionization era requires simulations of high dynamic range in mass and length scales to account for the key physical processes in the sources responsible for reionization at small length scales and also the details of large-scale effects such as evolution of underlying matter density fields, phase change of the IGM, and radiative feedback resulting from reionization. This requires us to have a simulation of a large cosmological volume and a high resolution.

One way to simulate reionization is to solve the cosmological radiative transfer equation along the path of every heating and ionizing photon \citep{iliev_2014_how_large, ricotti_2002_RT, thomas_2009_RT, Gnedin_2014_RT, ghara_2015_RT, Hirling_2024_PyC2Ray, Meriot_2025_RT, Ocvirk_2025_RT, Schwandt_2026_RT}. These 3D radiative transfer simulations help us study the evolution of the IGM with the progress of reionization, but are computationally very expensive. A computationally less expensive alternative is to use semi-numerical simulations of reionization \citep{Choudhury_2018_SCRIPT, Maity_2022_SCRIPT, Choudhury_2025_SCRIPT, Mondal_2017_reionyuga, Murray_2020_21cmFAST, Visbal_2012_21cmSPACE, Fialkov_2014_21cmSPACE, Dhandha_2025_21cmSPACE}. We focus on a semi-numerical technique based on the excursion set formalism \citep{Furlanetto_2004}, which compares the number of HI atoms to the number of HI ionizing photons present in a smoothing volume to generate the ionization maps \citep{majumdar_2014_sem-num, majumdar_2016_source-models}. After producing the ionization maps, we generate the 21-cm brightness temperature maps following the equation \ref{t_b} discussed in the next section. In this work, we have used the 21-cm differential brightness temperature maps that were generated using the semi-numerical simulation code \texttt{Sem-Num} \citep{TRC_inside-out, majumdar_2014_sem-num}. The semi-numerical simulation used for the analysis in this work is performed using the following steps:
\begin{itemize}
    \vspace{-0.25cm}
    \item Initially, the dark matter distribution is generated with an $N$-body gravity-only simulation at a set of redshifts ranging from $13.2$ to $7.2$.
    \vspace{-0.25cm}
    \item The Spherical overdensity halo finder algorithm
    is then employed on this dark matter particle distribution to identify the collapsed dark matter halos. These halos are considered to be the hosts of sources that emit ionizing photons.
    \vspace{-0.25cm}
    \item Once collapsed halos above a threshold halo mass hosting ionizing sources are identified, we model the spectrum of photons emitted from these halos and use the excursion set formalism to generate the ionization field, which is then converted to the 21-cm field.
\end{itemize}

By following the above procedure, we generate coeval cubes of size 500 $h^{-1}$ Mpc (714 Mpc) along each side on the comoving scale \citep{majumdar_2016_source-models}. The underlying dark matter distribution at each redshift was generated using a P$^3$M $N$-body simulation code \texttt{CUBEP$^3$M} \citep{harnois-deraps_2013_cubep3m} as part of the \texttt{PRACE4LOFAR} project. For the dark matter field, the authors of \citep{majumdar_2016_source-models} have used $6912^3$ particles of mass $4.0 \times 10^{7} M_\odot\ $ on a $13824^{3}$ mesh (comoving grid resolution of $0.052$ Mpc), and later these simulated matter particle positions and halo catalogs are interpolated on a $600^3$ grid (comoving grid resolution of $1.19$ Mpc). The parameters considered for the semi-numerical reionization simulation to generate the brightness temperature maps are the minimum halo mass that can host sources of ionizing photons ($M_{h,\rm min}$), mean free path of ionizing photons in the IGM ($R_{\rm mfp}$), and the number of ionizing photons entering the IGM per baryon ($N_{\rm ion}$). In these simulations, $M_{h,\rm min} = 2.09 \times 10^9 \, M_\odot\ $, whereas $R_{\rm mfp}$ and $N_{\rm ion}$ are model-dependent and are discussed in the following sections.

\subsection{Reionization source models and scenarios}
The differential brightness temperature of HI 21-cm signal is expressed as:
\begin{equation}\label{t_b}
\begin{split}
    \delta T_{b} (z) \approx 27x_{\rm HI}\left(1+\delta_{b}\right)\left(\frac{\Omega_{b}h^{2}}{0.023}\right)\left(\frac{0.15}{\Omega_{m}h^{2}}\frac{1+z}{10}\right)^{1/2}\left(1-\frac{T_{\rm{CMB}}(z)}{T_{s}}\right) \rm{mK}
\end{split}
\end{equation}
Where $x_{\rm{HI}}$ is the neutral hydrogen fraction, $\delta_{b}$ is the baryon overdensity, $T_{\rm{CMB}}$ and $T_s$ are the background CMB temperature and the spin temperature of HI, respectively. The 21-cm signal is either observed in emission or absorption depending on the relative value of $T_s$ with $T_{\rm{CMB}}$. During EoR, the spin temperature is strongly coupled to the kinetic temperature of the gas through the Wouthuysen-Field effect. The X-ray radiation from the first luminous sources is expected to rapidly heat the IGM well above the CMB temperature, thereby raising its spin temperature, during the early stages of reionization. In this work, we assume that the spin temperature of the IGM gas is much larger than the background CMB temperature ($T_s \gg T_{\rm{CMB}}$) and ignore the redshift-space distortions (RSDs) caused by the peculiar velocities of the gas.

The evolution of the topology of ionized bubbles depends on both the properties of sources emitting ionizing photons and the IGM characteristics. Thus, it is possible to generate topologically distinct ionization fields by varying the source characteristics and properties of the IGM. In this work, we consider four source properties resulting in five different reionization scenarios with different topological evolution of ionized bubbles. These scenarios were distinguished in coeval 21-cm maps in paper I, and we examine their distinguishability when the lightcone effect is implemented. In the following section, we briefly describe the source models considered in this work, detailed in \citep{majumdar_2016_source-models}.

\subsubsection{Reionization source models}
In this work, we consider four different source models based on the type and intensity of photons emitted by them. In the next section, we describe how these source models are combined to generate topologically distinct reionization scenarios.
\begin{itemize}
    \vspace{-0.2cm}
    \item \textbf{Ultraviolet photons (fiducial model)}: In this model, we assume that UV photons emitted from the galaxies residing in collapsed halos are responsible for reionization. It is also assumed that the number of ionizing photons emitted from a luminous galaxy follows the following relation:
     \begin{equation}
        N_{\gamma}(M_h) = N_{\rm ion} \frac{M_{h} \Omega_{\rm b}}{m_{\rm p} \Omega_{\rm m}} \, ,
        \label{eq:sm}
    \end{equation}
    where $m_p$ is the mass of a proton, $M_h$ is the mass of the halo that hosts the source, and $N_{\rm ion}$ is the number of ionizing photons entering the IGM per baryon.
    
    \vspace{-0.2cm}
    \item\textbf{Uniform Ionizing Background (UIB photons)}: The hard X-rays emitted by sources like X-ray binaries, AGNs, etc., are assumed to be the dominant cause of reionization in this model, and the number of photons emitted by these sources follows the relation \ref{eq:sm}. The X-rays emitted by these sources travel long distances before getting absorbed by the intervening IGM. The long mean free path of hard X-rays will result in a uniform ionizing photon distribution contributing to a uniform ionizing background. This results in an outside-in reionization scenario where low-density regions are ionized first and high-density regions later.

    \vspace{-0.2cm}
    \item\textbf{Soft X-ray photons (SXR)}: In this source model, soft X-ray photons generated using equation \ref{eq:sm} dominate reionization. These SXRs contribute to a uniform ionizing background limited by the mean free path of photons. The mean free path of SXR photons depends on the frequency and redshift of their origin and is determined by following the formalism used in \citep{mcquinn_2012_sxr}. In this work, we assume that all the SXR photons emitted by the sources have the same energy of 200 eV. The photons emitted are uniformly distributed in a spherical region around the source. The radius of this uniform distribution around the source is determined by the mean free path of SXR photons.

    \vspace{-0.2cm}
    \item\textbf{Power law mass dependent efficiency (PL)}: In this source model, we consider that the number of UV photons emitted by the ionizing source ($N_\gamma$) is proportional to the $n^{\rm th}$ power of the mass of the host dark matter halo ($M_h$).
    \begin{equation}
        N_{\gamma}(M_h) \propto M_h^n \, ,
        \label{eq:PL}
    \end{equation}
    In this work, we consider a power-law index of 3 in order to generate the differential brightness temperature maps for our analysis.
        
\end{itemize}
\subsubsection{Reionization Scenarios}
\label{sec:reionization_scenarios}

In this section, we discuss five different reionization scenarios generated by using the above-mentioned source models. The contribution of source models to different reionization scenarios is summarized in Table \ref{tab:reionization_scenarios}.

\begin{table}[htbp]
    \centering
    \begin{tabular}{|c|c|c|c|c|}
    \hline
        \textbf{Reionization Scenarios} &\textbf{UV}  &\textbf{UIB}  &\textbf{SXR}  &\textbf{PL,n} \\
        \hline
         Fiducial&100\% & - & - & 1.0\\
         \hline
         Clumping&100\% & - & - & 1.0 \\
         \hline
         PL($n=3$)&100\%  & - & - & 3.0\\
         \hline
         UIB dominated& 20\% &80\%  &-  &1.0 \\
         \hline
         UV+SXR+UIB&50\%  &10\%  &40\%  &1.0 \\
         \hline  
    \end{tabular}
    \caption{Contribution of different source models in our reionization scenarios.}
    \label{tab:reionization_scenarios}
\end{table}

The fiducial, clumping, and PL($n=3$) reionization scenarios assume that only UV photons emitted from the galaxies residing in collapsed halos of mass $\geq$ $2.09 \times 10^9 M_\odot$ contribute to reionization. In the clumping scenario, we account for the inhomogeneous spatial distribution of ionizing radiation sinks that remain self-shielded from the ionizing background, as discussed in \citep{TRC_inside-out}. This is implemented through a simple prescription in which regions with overdensities exceeding a critical threshold always remain strictly neutral, as suggested in \cite{Escude}. Depending upon the chosen threshold, this approach leads to an overestimation of self-shielded regions with high optical depth, like Lyman limit systems \citep{sargent_1989_lyman-limit, lanzetta_1991_lyman-limit}. However, it results in a topologically distinct 21-cm map compared to our fiducial scenario. For further details regarding the implementation of this reionization scenario, we refer the interested reader to Section 2.5 of \citep{TRC_inside-out}. In the PL ($n=3$) reionization scenario, high mass halos have a higher weightage of ionizing UV photon contribution (\ref{eq:PL}), resulting in the formation of much larger ionized bubbles around high mass halos relative to the fiducial scenario. In the other scenarios, like UIB and UV+SXR+UIB, we assume a varied contribution of ionizing photons from different source models as mentioned in Table \ref{tab:reionization_scenarios}. As the name suggests, the dominant contribution of hard X-rays to reionization leads to the formation of a uniform ionizing background in the UIB model. The UV+SXR+UIB model takes into account the contribution of three source models, making UV, Hard X-ray, and soft X-ray photons generate the ionization field.

The proportionality constant $N_{\rm ion}$ was tuned in the reionization scenarios considered (Table \ref{tab:reionization_scenarios}) such that they follow the same reionization history, i.e., they have the same mass-averaged ionized fraction at any given redshift. Since the observations of the future SKA-Low telescope have a lower resolution than 1.19 Mpc, we coarse-grid and downsample the brightness temperature maps of these reionization scenarios to the resolution of 2.38 Mpc following the procedure used in paper IIa. The downsampling is performed by keeping only one randomly chosen cell out of eight neighboring cells in the original simulated maps. By following this method, the physical volume of the map remains the same, but we reduce our map dimensions from $600^3$ grid units to $300^3$ grid units. Note that these reionization scenarios lead to distinct reionization topologies, and we study the evolution of their topology using percolation analysis.

\subsection{Simulating lightcone maps}

\begin{figure}[htbp]
    \centering
    \includegraphics[width=0.8\linewidth]{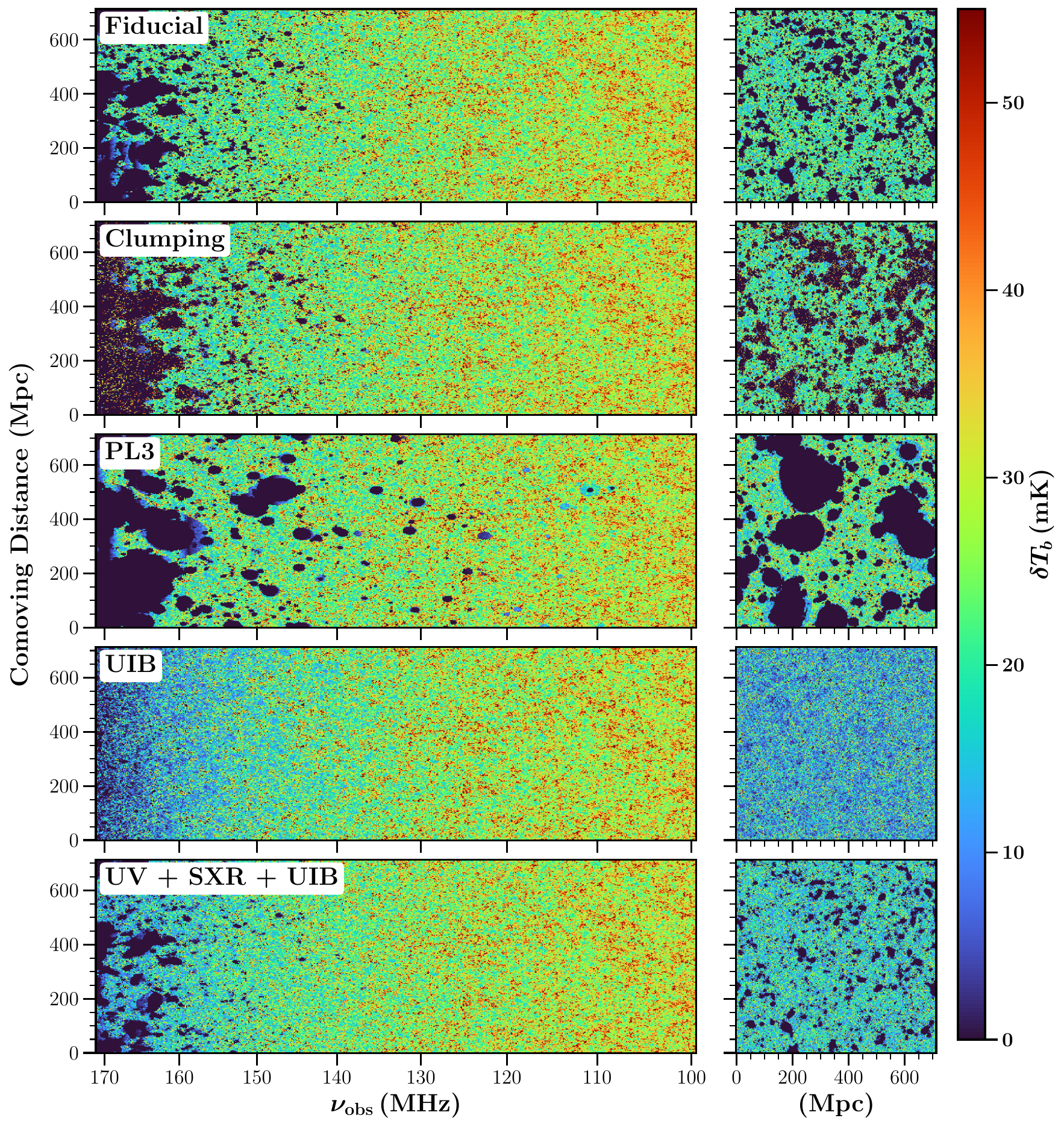}
    \caption{\textit{Left}: Corresponding slices cut along the line of sight of 21-cm brightness temperature lightcone maps for the fiducial, clumping, PL3, UIB, and UV+SXR+UIB reionization scenarios. The horizontal axis is the line of sight with reionization progressing from right to left. The observed frequencies $\nu_{\rm obs}$ correspond to the frequencies at which the 21-cm emission from the redshift of the slice can be observed. \textit{Right}: Corresponding slices of the same lightcone maps cut perpendicular to the line of sight at $z=7.91$.}
    \label{fig:lightcones}
\end{figure}
\begin{figure}[htbp]
    \centering
    \includegraphics[width=0.7\linewidth]{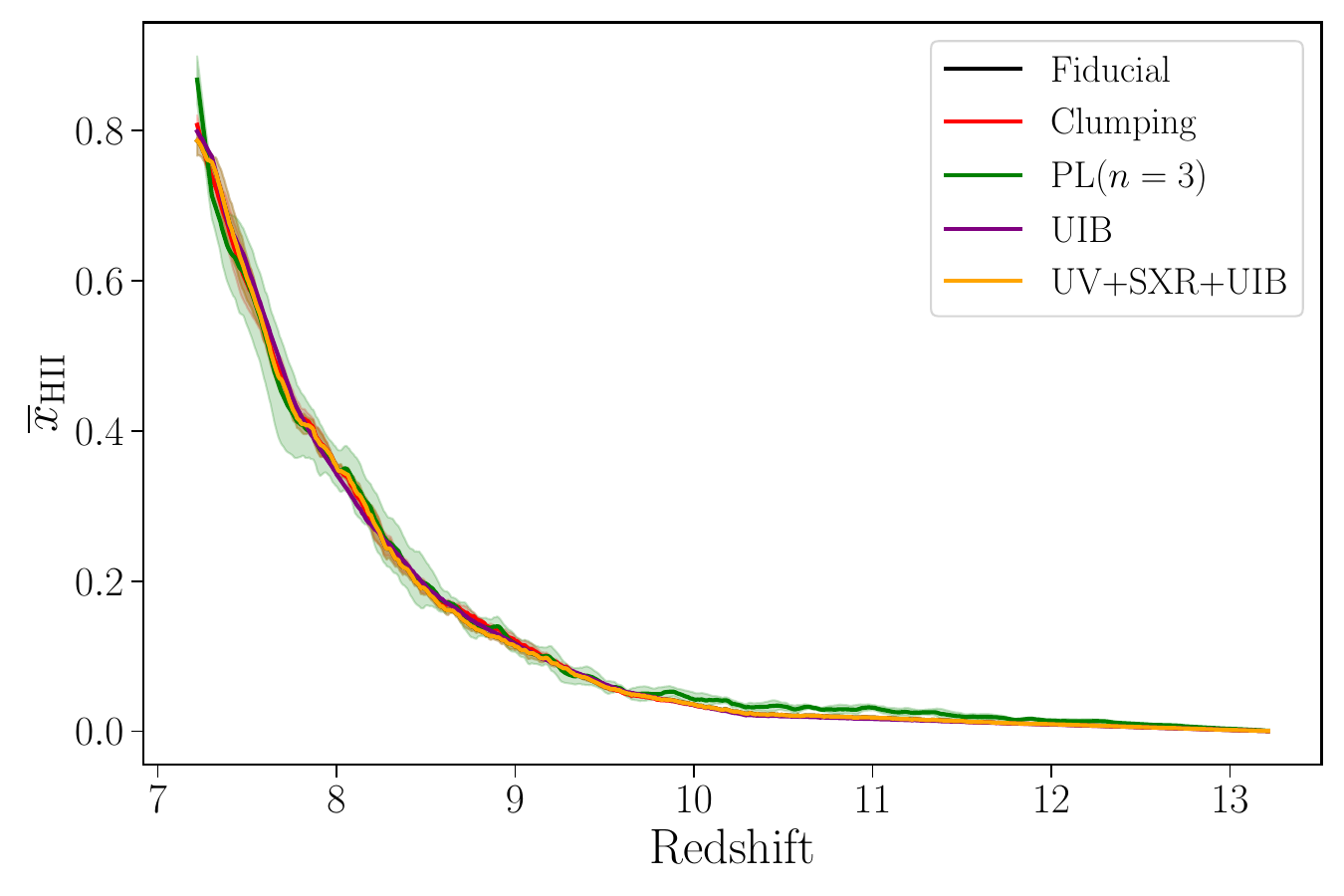}
    \caption{Evolution of the mass-averaged ionized fraction with redshift in 21-cm lightcones for the considered reionization scenarios. The shaded regions are $1\sigma$ errorbars computed from the variance across 18 different lightcone realizations.}
    \label{fig:reion_history}
\end{figure}
By following the procedure mentioned in the above sections, we generated coeval maps at redshifts ranging from $13.2$ to $7.2$ for each reionization scenario. We briefly outline the procedure for generating lightcone maps from coeval maps here. Note that the above-mentioned formalism is similar to the one employed in \citep{10.1111/j.1365-2966.2012.21293.x}.

\begin{itemize}
    \vspace{-0.25cm}
    \item We prepare a redshift list $Z$ with entries $z_{1}, z_{2}, z_{3}, ...., z_{N}$ ($z_{1} <z_{2} < z_{3}..... <z_{N}$) with $N = L/l$, where $L$ is co-moving distance between lowest ($z_1$) and highest ($z_N$) redshifts in the generated coeval boxes and $l$ is the grid resolution of the coeval maps.
    
    \vspace{-0.25cm}
    \item To construct the $p^{\rm th}$ slice of a lightcone ($z_p$) using the coeval cubes containing $M$ slices, we first compute the remainder $q$ of the integer division $\frac{p}{M}$. We then determine the two neighboring redshifts $z_{l}$ and $z_{l+1}$ at which coeval cubes are available such that $z_{l} \le z_{p} \le z_{l+1}$. The $p^{\rm th}$ slice of the lightcone is computed by linear interpolation of the $q^{\rm th}$ slices of $z_{l}$ and $z_{l+1}$.
\end{itemize}

Once the lightcone maps are generated following the above-mentioned approach for each reionization scenario, we relabel the line-of-sight axis with the observed frequency ($\nu_{\rm obs}$) in these maps using

\begin{equation}
    \nu_{\rm obs}=\frac{1420}{1+z} \, \rm MHz \, ,
    \label{eq:of}
\end{equation}
where $z$ is the redshift of the slice we consider and $1420$ MHz is the rest-frame frequency of 21-cm line emission. In order to account for the cosmic variance in our analysis, we generated 18 realizations of lightcone maps for each reionization scenario. This is done by varying the starting position within the coeval maps by making periodic shifts and also rotating the coeval boxes along the two other axes perpendicular to the initially considered line-of-sight axis, and following the procedure discussed above to generate the lightcones. We also generate lightcones of ionized fraction and matter density field, which are used to generate the 21-cm brightness temperature maps, and compute the mass-averaged ionized fraction $\bar{x}_{\rm HII}$ in each slice along the LOS. The redshift evolution of $\bar{x}_{\rm HII}$ remains the same for all the reionization scenarios considered, as shown in Figure \ref{fig:reion_history}. The shaded regions in all figures of this paper represent the $1 \sigma$ scatter computed across 18 realizations of lightcones unless otherwise specified.

\subsection{Instrumental effects and systematics}
\label{sec:SKA_effects}
Radio interferometric observations lack absolute flux calibration due to the absence of zero-length baselines, i.e., the $k_{\perp}=0$ wavenumber. To mimic this effect, we subtract the mean from each slice of the lightcones of the fiducial reionization scenario. As a result, the completely ionized regions in our maps, which correspond to $\delta T_b =0$, get mapped to the minima of the mean-subtracted map in each slice. In addition to this, a radio interferometer produces images after convolving the sky brightness with the beam of the telescope, resulting in a smoothing effect that degrades the image resolution.
In order to mimic this effect, we employ 2D Gaussian kernels of varying smoothing scales starting from 7.14 Mpc to 14.28 Mpc along the sky plane of the simulation volume, as in paper IIa.

Along with the above effects, we also include SKA-Low-like thermal noise, approximated by a Gaussian random field of RMS given by the radiometer equation \citep{taylor_synthesis_1999}:
\begin{equation}
\sigma_n = \frac{S_{\rm D} \times {\rm SEFD}}
{\eta_{\rm s} \sqrt{0.5 n_{\rm ant}(n_{\rm ant}-1)\, n_{\rm pol}\,
\Delta\nu\, \Delta\tau}},
\label{eq:sigma_n}
\end{equation}
where $S_{\rm D} = 2$ is a degradation factor with respect to the natural array
sensitivity, $\eta_{\rm s} = 0.9$ is the system efficiency, $n_{\rm ant} = 512$ is the number of SKA-Low stations in the AA4 configuration \citep{seethapuram_sridhar_2025_AA4}, $n_{\rm pol} = 2$ is the number of polarizations,
$\Delta\nu$ is the effective spectral resolution, and $\Delta\tau = 5000$~h is the total integration time. The system equivalent flux density is
\begin{equation}
{\rm SEFD}(\nu) = \frac{2 k_B}{A_{\rm eff}/T_{\rm sys}(\nu)},
\label{eq:sefd}
\end{equation}
where $k_B$ is the Boltzmann constant, $A_{\rm eff}$ is the effective area of the telescope and $T_{\rm sys}$ is  system temperature. To estimate the SEFD, we consider the SKA-low sensitivity values from the SKA-Low baseline design \citep{braun_anticipated_2019} \footnote{\href{https://www.skao.int/sites/default/files/documents/SKAO-TEL-0000818-V2_SKA1_Science_Performance.pdf}{https://www.skao.int/sites/default/files/documents/SKAO-TEL-0000818-V2\_SKA1\_Science\_Performance.pdf}} for the AA4 configuration and interpolate them linearly onto the observed frequency channels of the lightcone. As the system temperature of the telescope is dominated by the Galactic synchrotron emission, which scales as a steep inverse power law with frequency ($T_{\rm sky} \propto \nu^{-2.55}$), the SEFD and consequently the RMS of the noise increase significantly at lower observed frequencies.
The RMS of noise is estimated in Jy~beam$^{-1}$ and is converted to brightness
temperature units using the Rayleigh-Jeans approximation,
\begin{equation}
\sigma_T = \frac{\lambda^2}{2 k_{\rm B} \Omega}\, \sigma_n,
\label{eq:sigma_T}
\end{equation}
where $\lambda$ is the redshifted wavelength of the 21-cm line and $\Omega$ is the solid
angle subtended by the resolution element. For a Gaussian beam,
\begin{equation}
\Omega = \frac{\pi \theta^2}{4 \ln 2},
\label{eq:omega}
\end{equation}
where $\theta$ is the angular resolution. The angular scale corresponding to a comoving
scale $L$ at redshift $z$ is $\theta = L / D_{\rm A}(z)$, with $D_{\rm A}(z)$ the angular diameter distance. To balance the need for a fine spectral resolution against low thermal noise, we downsample the lightcone along the frequency axis by a factor of two, averaging two adjacent channels. This frequency averaging lowers the noise RMS per channel relative to the original resolution. We add Gaussian random noise with RMS given by $\sigma_T(\nu)$ to each frequency slice of the mean-subtracted, beam-smoothed lightcone. The noise is drawn independently for each realization, so that the $1\sigma$ scatter across the 18 realizations discussed in Section \ref{sec: noise_results} includes both cosmic variance and noise variance.

\begin{figure}[!htbp]
    \centering
    \includegraphics[width=\linewidth]{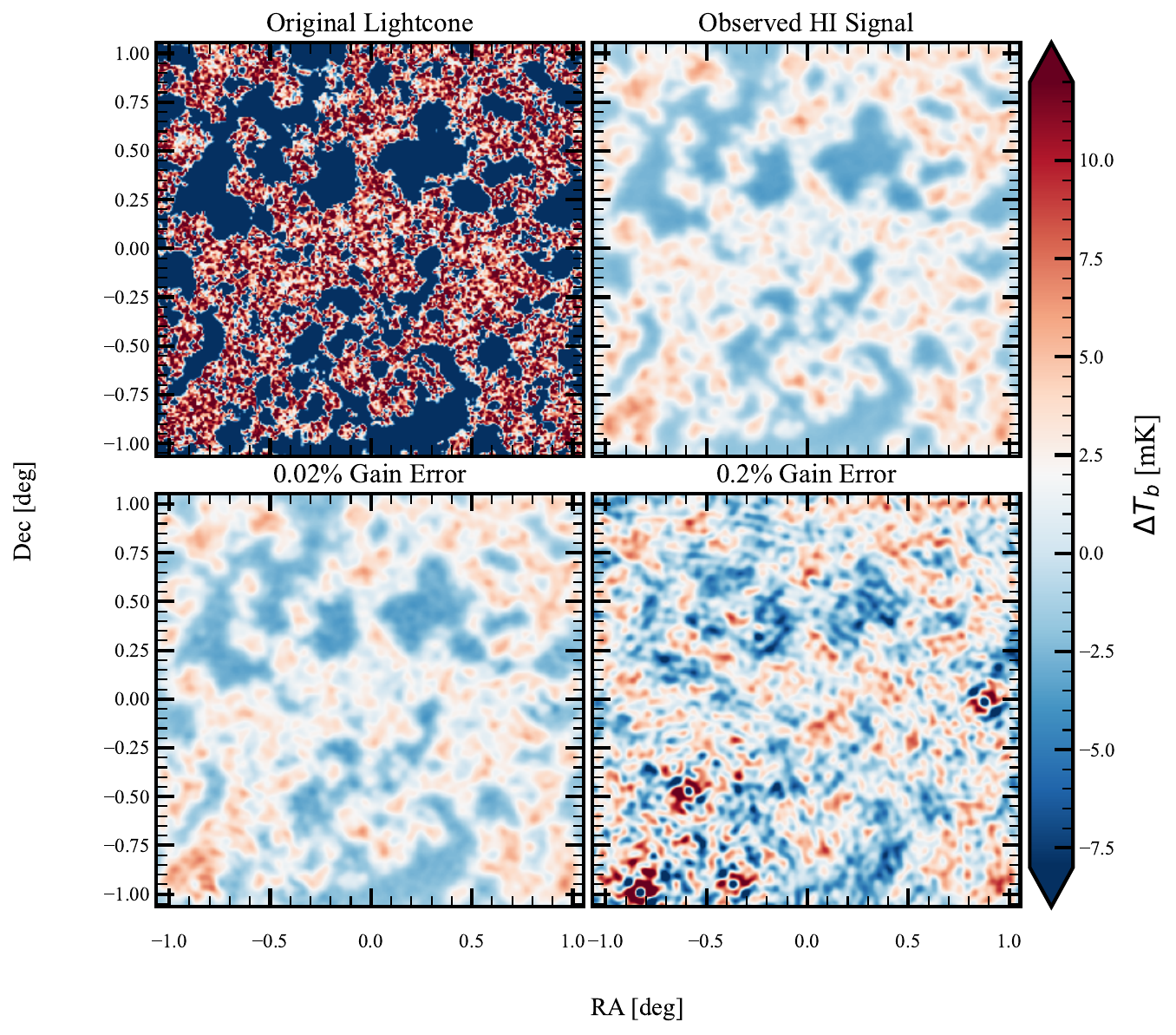}
    \caption{Slices of the observed 21-cm lightcone maps at 165.87 MHz. From left to right in the top panel: the lightcone map without instrumental effects, the map including synthesized beam effect but without gain calibration errors. From left to right in the bottom panel: the map including synthesized beam effect with  $\sim 0.02\%$ gain calibration error and the map including synthesized beam effect with  $\sim 0.2\%$ gain calibration error.}
    \label{fig:ska-low_observed_map}
\end{figure}

In order to produce realistic mock 21-cm observations, we simulate synthetic radio observations tailored to the characteristics of the upcoming SKA-Low AA4 configuration. Specifically, we use our in-house developed end-to-end simulation pipeline, 21cmE2E\footnote{\url{https://gitlab.com/samit-pal/21cme2e.git}} (built upon the \texttt{OSKAR}\footnote{\href{https://github.com/OxfordSKA/OSKAR.git}{https://github.com/OxfordSKA/OSKAR.git}} package \citep{Dulwich2020zndo...3758491D}), which incorporates realistic sky models, telescope beam responses, and instrumental and environmental effects \citep{10.1093/mnras/stac1994, Pal_2025, samit_lcs}. 
In this work, we investigate the impact of antenna-based gain calibration errors on the  21-cm lightcone maps. To construct the observational mock data, we transform the physical lightcone into an observed differential brightness temperature cube $\Delta T_b(\theta_x, \theta_y, \nu_{\text{obs}})$ with FoV of $2.13^{\circ} \times 2.13^{\circ}$. This transformation maps comoving spatial coordinates $(x, y, z)$ in Mpc to observational sky angular coordinates $(\theta_x, \theta_y)$ in degrees and line-of-sight observed frequencies $\nu_{\text{obs}}$ in $\text{MHz}$. The coordinate conversion and gridding are implemented using the \texttt{Tools21cm}\footnote{\href{https://github.com/sambit-giri/tools21cm.git}{https://github.com/sambit-giri/tools21cm.git}} python package \citep{giri2020tools21cm}. The input sky model combines the cosmological 21\,cm signal with extragalactic point sources drawn from a composite of the GaLactic and Extragalactic All-sky MWA (GLEAM) \citep{Wayth2015, Hurley-Walker2017MNRAS.464.1146H} and Long Baseline Epoch of Reionization Survey (LoBES) catalogs \citep{Lynch2021}. We simulate a 4-hour tracking observation spanning an hour-angle range from $-2\mathrm{h}$ to $+2\mathrm{h}$, centered at $(\alpha, \delta) = (0\mathrm{h}, -30^{\circ})$. The simulation covers a frequency range of $130.87 - 172.87\,\text{MHz}$ with a $140\,\text{kHz}$ channel resolution. Each visibility sample is integrated over 120\,seconds,  yielding 120\,time steps for the full observation, chosen to reduce computational cost. To incorporate gain calibration errors, we followed the methodology outlined in paper IIb, with final post-calibration and post-averaging standard deviation levels of 0.02\% and 0.2\% with respect to deep 1000 hours of observation, representing optimistic and pessimistic scenarios, respectively. Because of the high computational cost of these end-to-end simulations, we generate these gain-calibrated mock datasets for only one lightcone realization. The instrumental effects and systematics discussed in this section are applied only to the fiducial reionization scenario in this work. Figure~\ref{fig:ska-low_observed_map} shows a comparison of one such slice of the observed 21-cm maps at different stages of the observation simulation. The maps illustrate the impact of the synthesized beam and calibration errors introduced using the OSKAR simulation.

\section{Methodology}
\label{sec:methodology}

\subsection{Percolation analysis}
The first luminous sources, once formed ($z \approx 15$), emit X-ray and UV photons, which heat and subsequently ionize the neutral Hydrogen present in the intergalactic medium (IGM). As this process of reionization progresses, new ionized regions start forming, and existing ionized regions grow and overlap with each other. As time progresses, these ionized regions get interconnected to form a single large ionized region extending throughout the IGM. This phase transition, resulting in the formation of a single LIR, is called the percolation transition \citep{10.1093/mnras/sty714, 10.1093/mnras/stz532, Pathak_2022, Dasgupta_2023, samit_lcs}.

In this work, we use the Largest Cluster Statistic (LCS) \citep{1993ApJ...413...48K, Bharadwaj_2000, 10.1093/mnras/sty714, Pathak_2022, Dasgupta_2023, samit_lcs} to track the percolation process in lightcone maps and determine the timing of the percolation transition in different reionization scenarios. LCS gives a measure of the fraction of ionized volume present in the LIR and is defined as follows:

\begin{equation}
    \text{LCS}=\frac{\text {Volume of the largest ionized region}}{\text{Total volume of all the ionized regions}} \, .
    \label{LCS}
\end{equation}
The Filling Factor (FF) is another statistic used to probe the ionization state of the IGM. FF gives a measure of the fraction of the ionized volume present in the simulation volume considered and is defined as follows:

\begin{equation}
    \text{FF}=\frac{\text {Total ionized volume}}{\text{Volume of the simulation box}} \, .
    \label{FF}
\end{equation}

After percolation, the entire ionized volume is carried by the LIR, extending from one end of the simulation box to the other, resulting in an LCS of 1. The FF is the volume-averaged ionized fraction, which helps us determine the evolution of the ionization state of the IGM during reionization.

To compute the LCS and FF in a given tomographic volume, we use the \texttt{SURFGEN2} algorithm, whose detailed description can be found in \citep{10.1093/mnras/sty714, 10.1093/mnras/stz532, sheth_2003_surfgen}. \texttt{SURFGEN2} binarizes the given tomographic volume based on a threshold value and identifies the ionized regions by the Friends-of-Friends (FoF) algorithm. Once the ionized regions are identified, their Shapefinders \citep{Sahni_1998_shapefinders} are computed using the \texttt{Marching Cubes 33} algorithm \citep{chernyaev_1995_marchingcube33}, and the values of LCS and FF are also calculated. Note that \texttt{SURFGEN2} counts only fully ionized cells while computing cluster statistics. 

\subsection{Calculating LCS in lightcone maps}
\label{sec:LCS_on_lightcones}

\begin{figure}[htbp]
    \centering
    \includegraphics[width=0.49\textwidth]{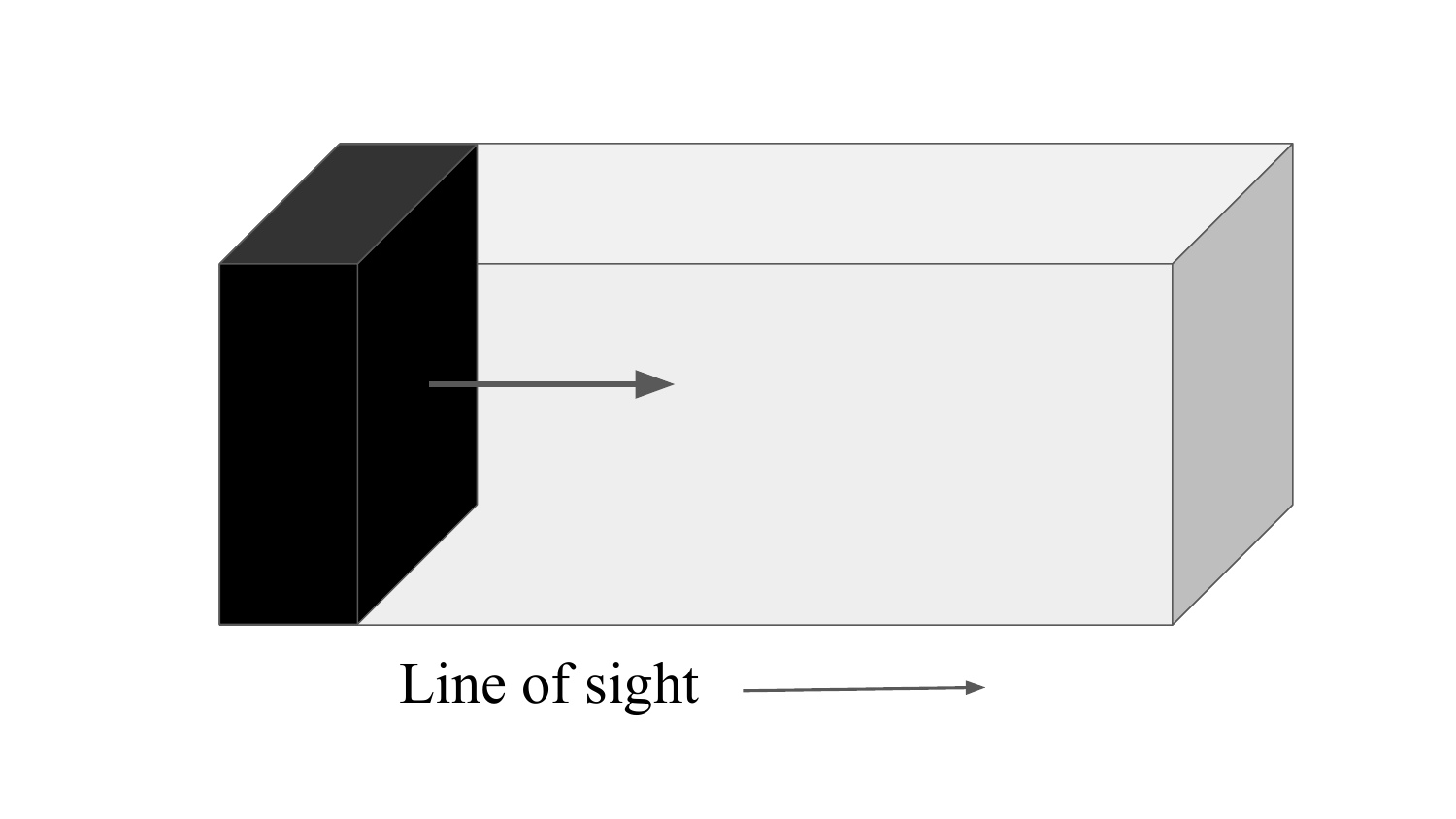}
    \hfill
    \includegraphics[width=0.49\textwidth]{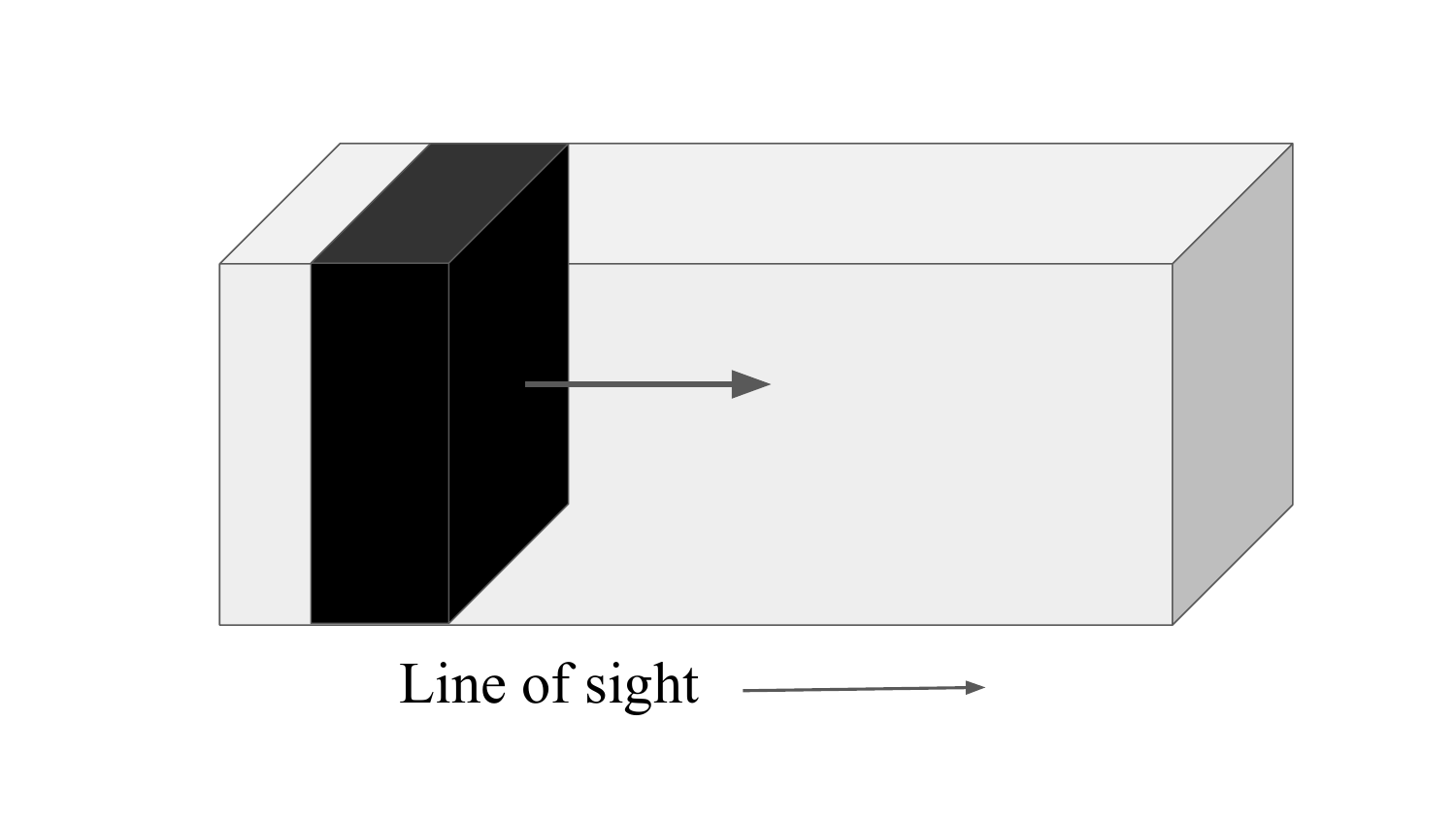}
    \caption{Illustration of the moving volume approach with the lightcone represented as a grey solid box, and the moving volume represented as a black solid box. The figure on the left-hand side represents the first iteration, and the figure on the right-hand side represents the second iteration, where the moving volume is moved by one grid unit.}
    \label{Moving-volume figure}
\end{figure}

When \texttt{SURFGEN2} is employed on a lightcone map, it outputs a single value of LCS corresponding to that map, with which one cannot understand the evolution of LCS with observed frequency and hence redshift. In order to understand the evolution of the cluster statistics, we divide lightcone maps into subvolumes by using a moving volume approach as illustrated in Figure \ref{Moving-volume figure}. We choose a subvolume (called a moving volume) of fixed frequency width along the line of sight and move it along the line of sight by one simulation grid unit at a time, starting from the highest observed frequency. We then compute cluster statistics using \texttt{SURFGEN2} in the subvolume corresponding to the location of the moving volume at each step. The cluster statistics are then labeled with the mean observed frequency at that step. By following this approach, we compute the evolution of cluster statistics in a lightcone map as discussed in section \ref{sec:results}.

\section{Results}
\label{sec:results}

\subsection{Percolation analysis on coeval maps}
The percolation behavior of ionized bubbles in coeval maps has been extensively studied in earlier works, including papers I, IIa, and IIb. In this subsection, we summarize the results from these works before discussing the results from this paper.

\begin{itemize}
    \vspace{-0.25cm}
    \item A sharp increase in the value of LCS has been observed at the onset of percolation in all the reionization scenarios considered from Table \ref{tab:reionization_scenarios}. The redshift (or observed frequency) at which this percolation transition occurs depends on the reionization scenario, indicating a strong dependence on the ionized region topology and can be used to distinguish between different reionization source models \citep{Pathak_2022}.
    
    \vspace{-0.25cm}
    \item Among the reionization scenarios considered, the ionized bubbles in coeval maps percolate at the earliest stage of reionization for the clumping model and at the most advanced stage for the UIB model. Additionally, the number of interconnections among the ionized regions, determined by genus, is much higher in the clumping scenario due to the formation of neutral islands that tunnel through the ionized regions due to density-dependent recombination \citep{Pathak_2022}.
    
    \vspace{-0.25cm}
    \item  The system noise and synthesized beam of SKA-Low introduce a bias in determining the onset of percolation. A robust recovery of the reionization history requires a minimum of 2000 hrs of observation to suppress the thermal noise, and a post-calibration antenna-based gain error tolerance of approximately $0.02 \%$ to prevent artificial fragmentation of the LIR \citep{Dasgupta_2023, samit_lcs}.
    
\end{itemize}

The analyses summarized above were primarily conducted using coeval boxes. In this work, we extend this study by applying percolation analysis to lightcone maps, presented in the subsequent sections.

\subsection{Percolation analysis on fiducial lightcone map}
\begin{figure}[htbp]
    \centering
    \begin{minipage}[t]{0.49\textwidth}
        \centering
        \includegraphics[width=\linewidth]{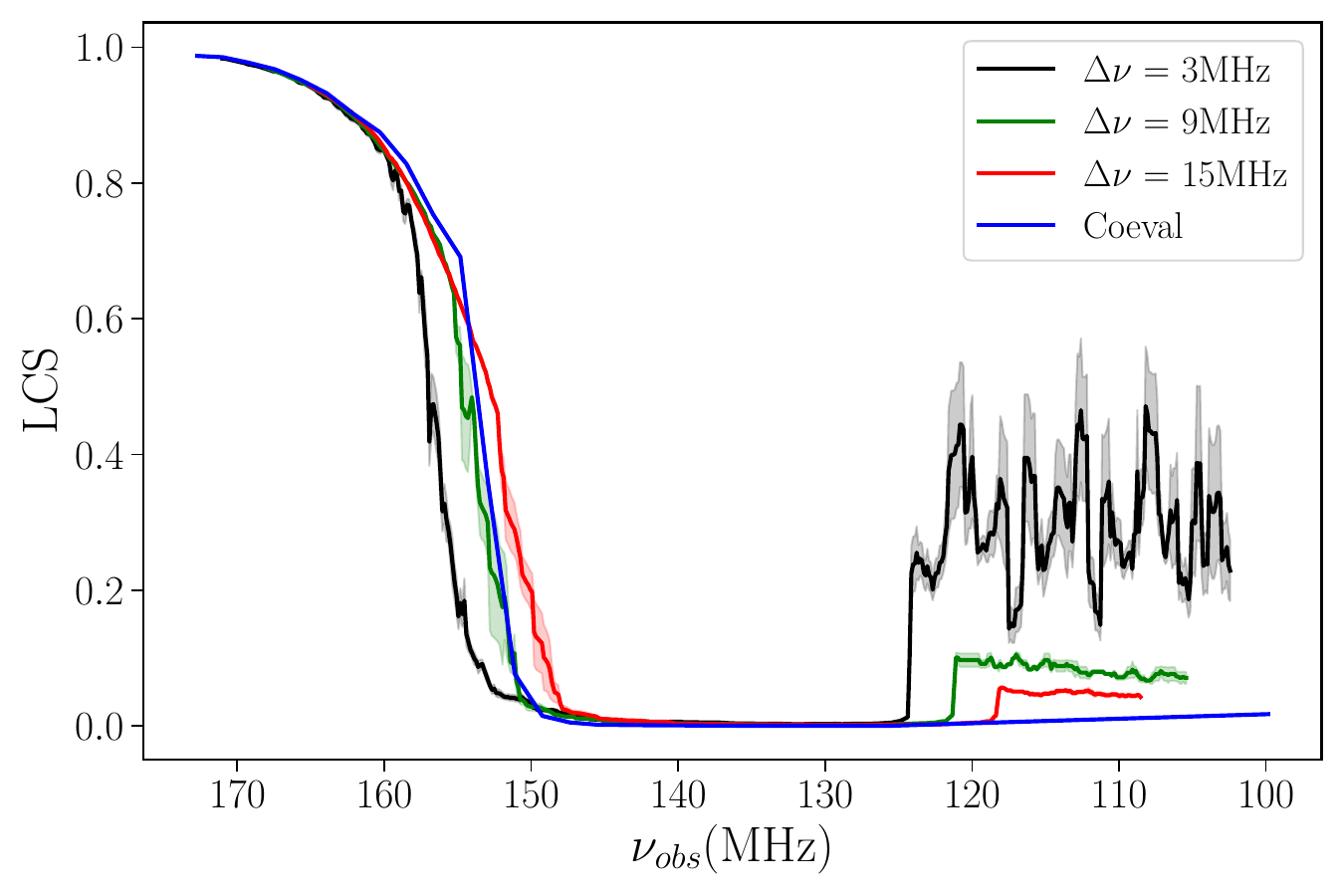}
    \end{minipage}
    \hfill
    \begin{minipage}[t]{0.49\textwidth}
        \centering
        \includegraphics[width=\linewidth]{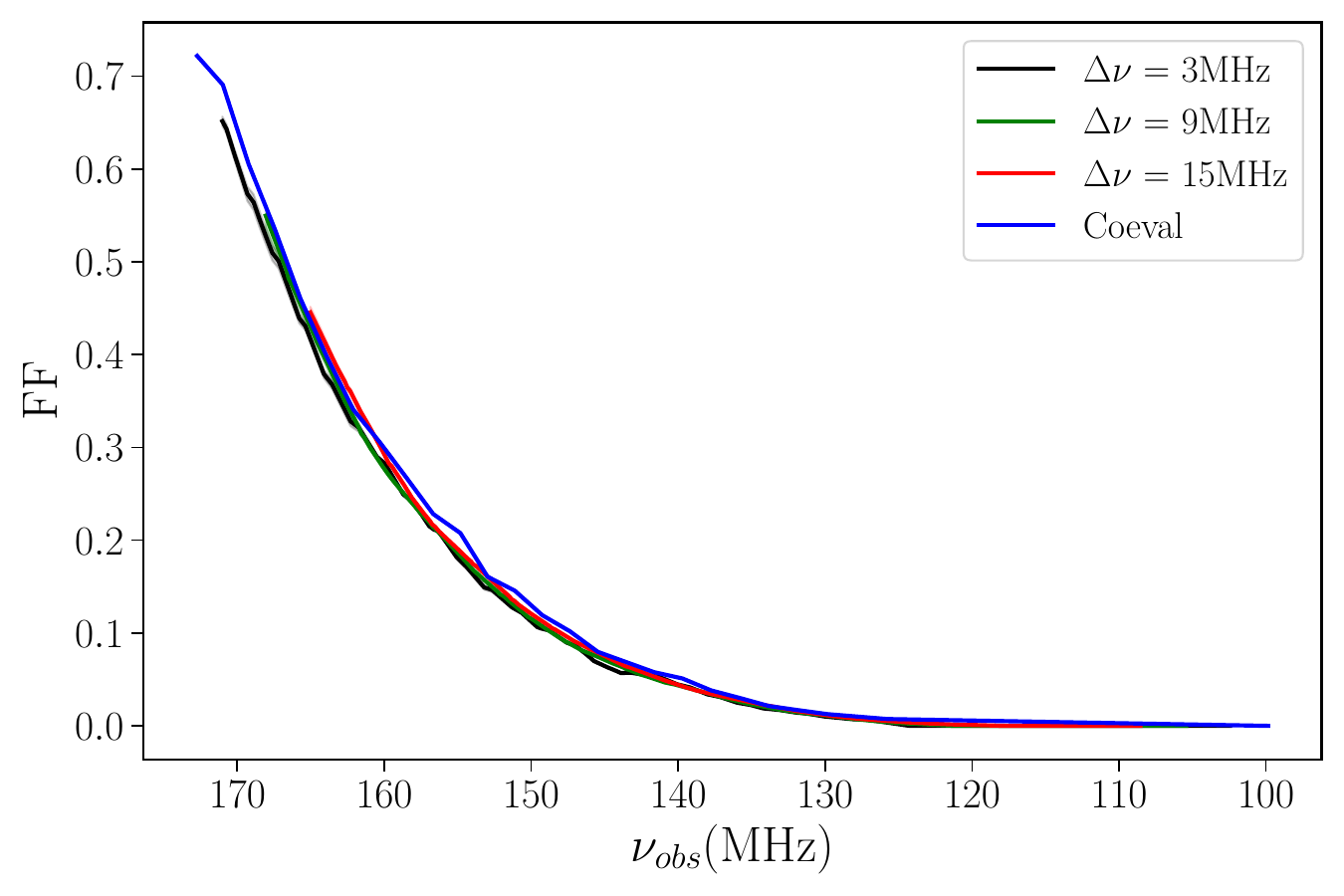}
    \end{minipage}
                                                                 
\caption{Evolution of the LCS (left) and FF (right) of moving volumes of different spectral bandwidths $\Delta \nu$ with their observed central frequency $\nu_{\rm obs}$ for the fiducial 21-cm lightcone.}
    \label{fig:lcs_fidu}
\end{figure}

The percolation transition, characterized by a steep rise in the LCS, can be estimated by tracking the evolution of the LIR within a simulation box. In this section, we employ a moving volume approach to track the evolution of the LIR in fiducial lightcone tomographic maps and compare it with its coeval counterpart to understand the bias introduced in the timing of percolation when the lightcone effect is considered. The subvolumes generated using the moving volume approach in this work are chunks of the lightcone map whose depth depends on the frequency width chosen to generate them. To understand the impact of the size of a moving volume, we perform this analysis considering spectral bandwidths of 3 MHz, 9 MHz, and 15 MHz along the line of sight in lightcone maps. The LCS and FF are calculated in all these subvolumes following the procedure outlined in section \ref{sec:LCS_on_lightcones}. Additionally, we compute the LCS and FF of the coeval boxes of the fiducial scenario, and compare their redshift evolution with that of the moving volumes along its lightcone. In Figure \ref{fig:lcs_fidu}, we present the evolution of LCS with observed frequency in the fiducial lightcone map for different moving volume spectral widths along with its coeval counterpart. 

We observe that each moving volume size exhibits a characteristic shape in the evolution of LCS. As the bandwidth of the moving volume increases, the onset of percolation begins at an earlier stage of reionization (lower observed frequency). Additionally, LCS assumes a relatively high value for smaller moving volume bandwidths in lightcone maps at lower observed frequencies (beginning of reionization) when compared to the LCS in coeval maps at that stage. This high value of LCS does not indicate the percolation of ionized bubbles, which is observed at later stages. Instead, it is attributed to the formation of small ionized bubbles at the beginning of reionization, among which the largest bubble occupies a very large fraction of the total ionized volume in the subvolume chosen at that frequency for the analysis. Lightcone slices within the frequency range of 100 MHz ($z = 13.2$) to 125.88 MHz ($z = 10.28$) are generated by linearly interpolating between the coeval snapshots at these frequencies, with no intermediate snapshots available. The interpolated cells will be fully ionized only if the corresponding cells are fully ionized in both snapshots, and the ionized regions in this range are essentially a few small isolated bubbles in the coeval snapshot corresponding to the 100 MHz observed frequency. Thus, when the high-frequency edge of the moving volume drops below the observed frequency of 125.88 MHz, an abrupt increase in the value of LCS is observed. Furthermore, the 3 MHz moving volume bandwidth captures a narrow comoving thickness along the LOS. This limited depth lacks a sufficient number of ionized regions to compute a statistically stable evolution of LCS, resulting in a high degree of fluctuations as shown in the black curve of the left panel in Figure \ref{fig:lcs_fidu}. As we increase the bandwidth of the moving volume, the subvolume chosen at that frequency accommodates more ionized bubbles in the chunk, thereby reducing the value of LCS. This high value of LCS at lower observed frequencies is not observed in full coeval maps, as the large size of these maps accounts for a large number of ionized bubbles, resulting in a decrease in the fraction of ionized volume occupied by the LIR. 

As the moving volume moves towards later stages of reionization, the formation of new ionized bubbles in the subvolumes used to compute the percolation statistics leads to an increase in the ionized volume. This results in a steep fall in LCS for low moving volume sizes. At the onset of percolation, ionized bubbles begin to coalesce, and the volume fraction occupied by the LIR increases rapidly, approaching unity. FF increases with observed frequency for all moving volume sizes considered, indicating an increase in ionization volume or the progression of reionization with time within the subvolumes of the lightcone map. We find that the FF estimated using the moving volume approach in lightcone maps has identical evolution to that of the FF in coeval maps. This evolution of FF for all moving volume sizes is observed to be consistent for all the reionization scenarios we discuss in the next section. 

\begin{figure}[htbp]
    \centering
    \begin{minipage}[t]{0.49\textwidth}
        \centering
        \includegraphics[width=\linewidth]{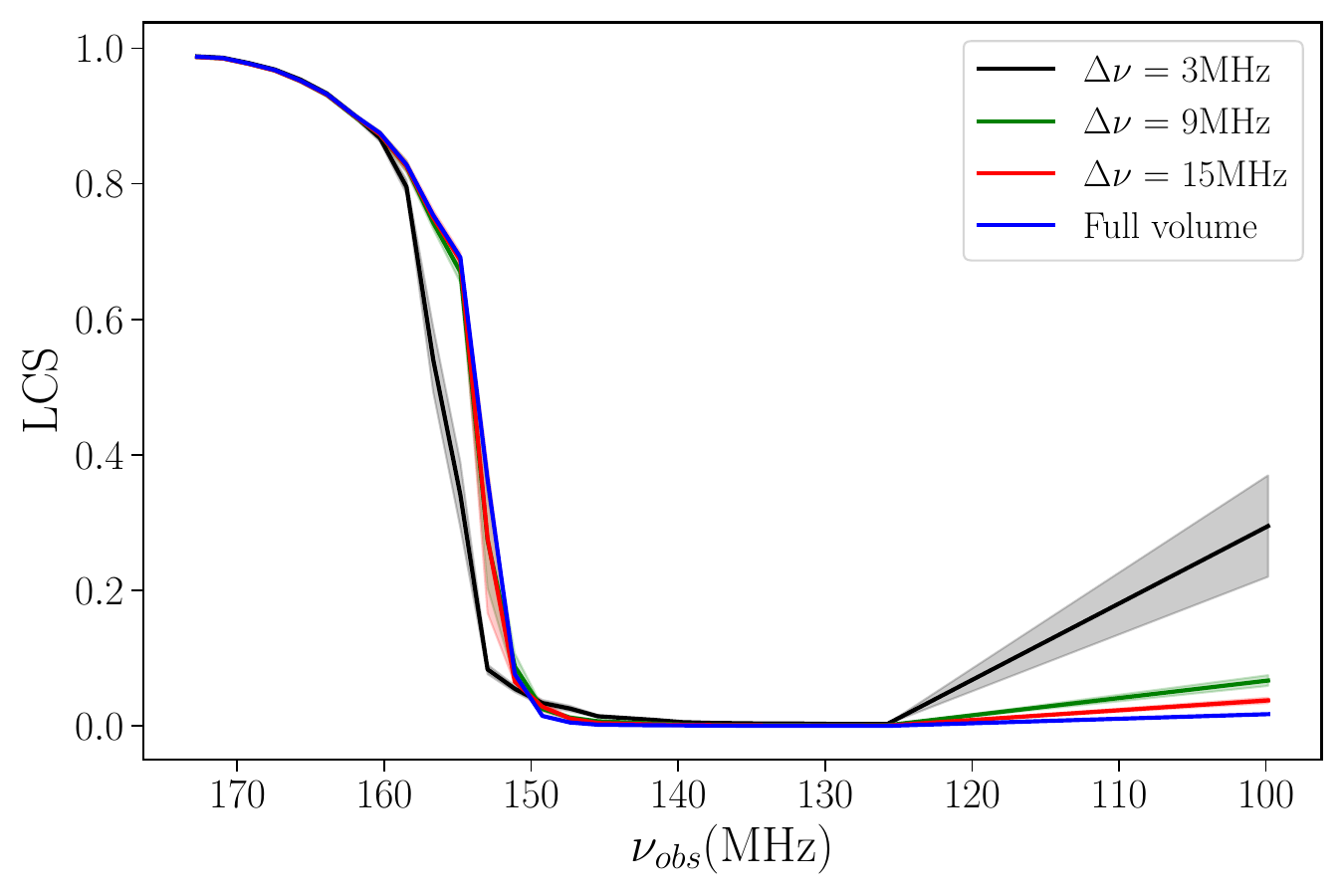}
    \end{minipage}
    \hfill
    \begin{minipage}[t]{0.49\textwidth}
        \centering
        \includegraphics[width=\linewidth]{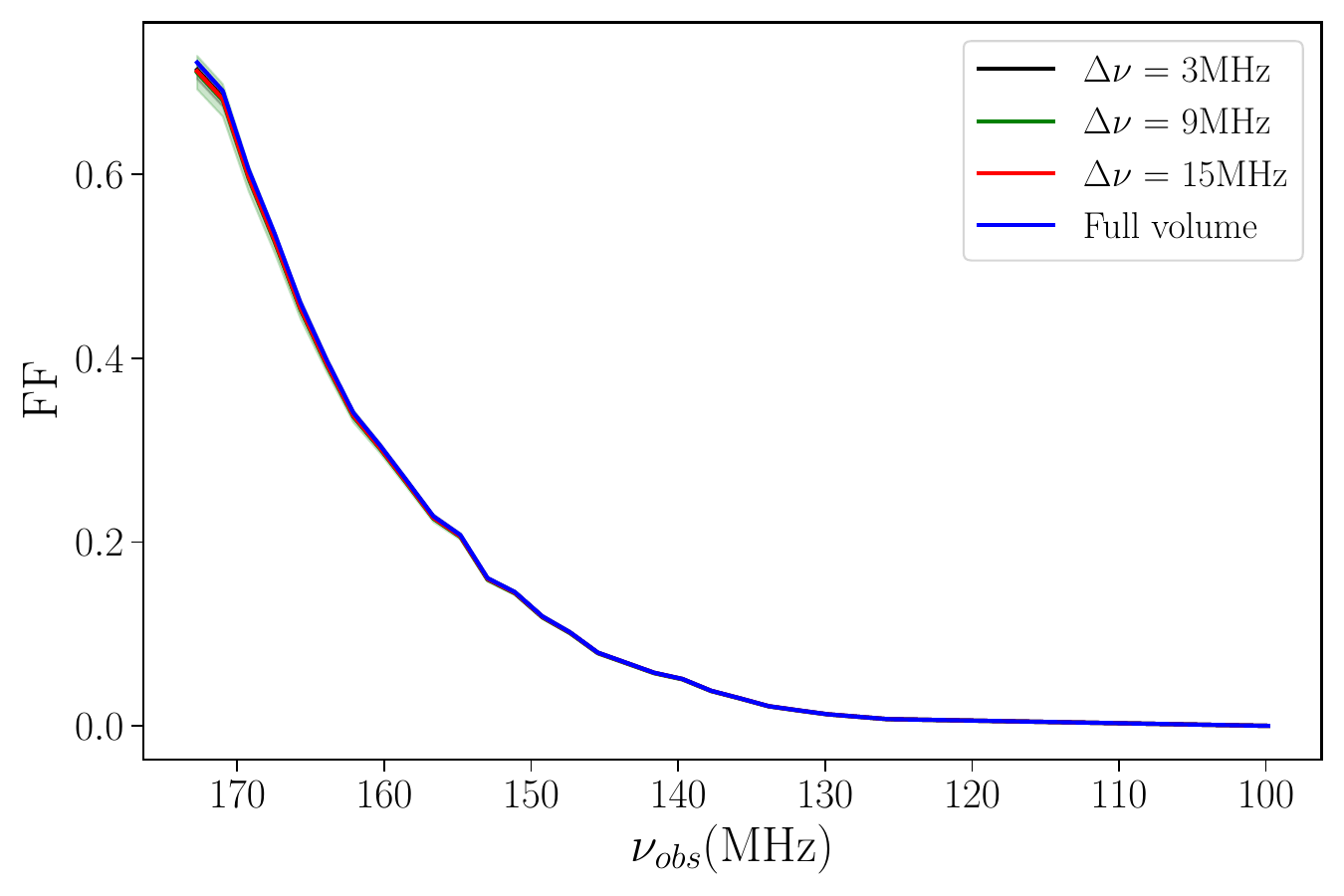}
    \end{minipage}
                                                                 
    \caption{Evolution of the LCS (left) and FF (right) of chunks of fiducial 21-cm coeval maps with comoving depth corresponding to different spectral bandwidths $\Delta \nu$ with their observed central frequency $\nu_{\rm obs}$. The shaded regions here represent the 1$\sigma$ scatter derived from non-overlapping coeval chunks of a given $\Delta \nu$, with the number of chunks considered depending on the chosen bandwidth.}
    \label{fig:lcs_fidu_coeval}
\end{figure}

Additionally, we divide our coeval maps into non-overlapping chunks of spectral bandwidths 3 MHz, 9 MHz, and 15 MHz and compute the percolation statistics, whose evolution is shown in Figure \ref{fig:lcs_fidu_coeval}. We perform this analysis only in the fiducial reionization scenario to understand the extent of bias introduced only through the lightcone effect rather than the finite width of the chunk.  We find that for both 9 MHz and 15 MHz sizes, the onset of percolation happens at the same observed frequency as that of the full coeval map. In the case of lightcone maps, the onset of percolation for the moving volume size of 9 MHz coincides with that of the full coeval map, whereas for the moving volume size of 15 MHz, the onset appears to happen at the earlier stages of reionization due to redshift evolution. This indicates that the moving volume size of 9 MHz is optimal to balance between minimizing the redshift evolution and maintaining the statistical significance in determining the onset of percolation in our fiducial reionization scenario.

\subsection{Evolution of LIR in different reionization scenarios}

\begin{figure}[htbp]
    \centering
    \begin{minipage}[t]{0.49\textwidth}
        \centering
        \includegraphics[width=\linewidth]{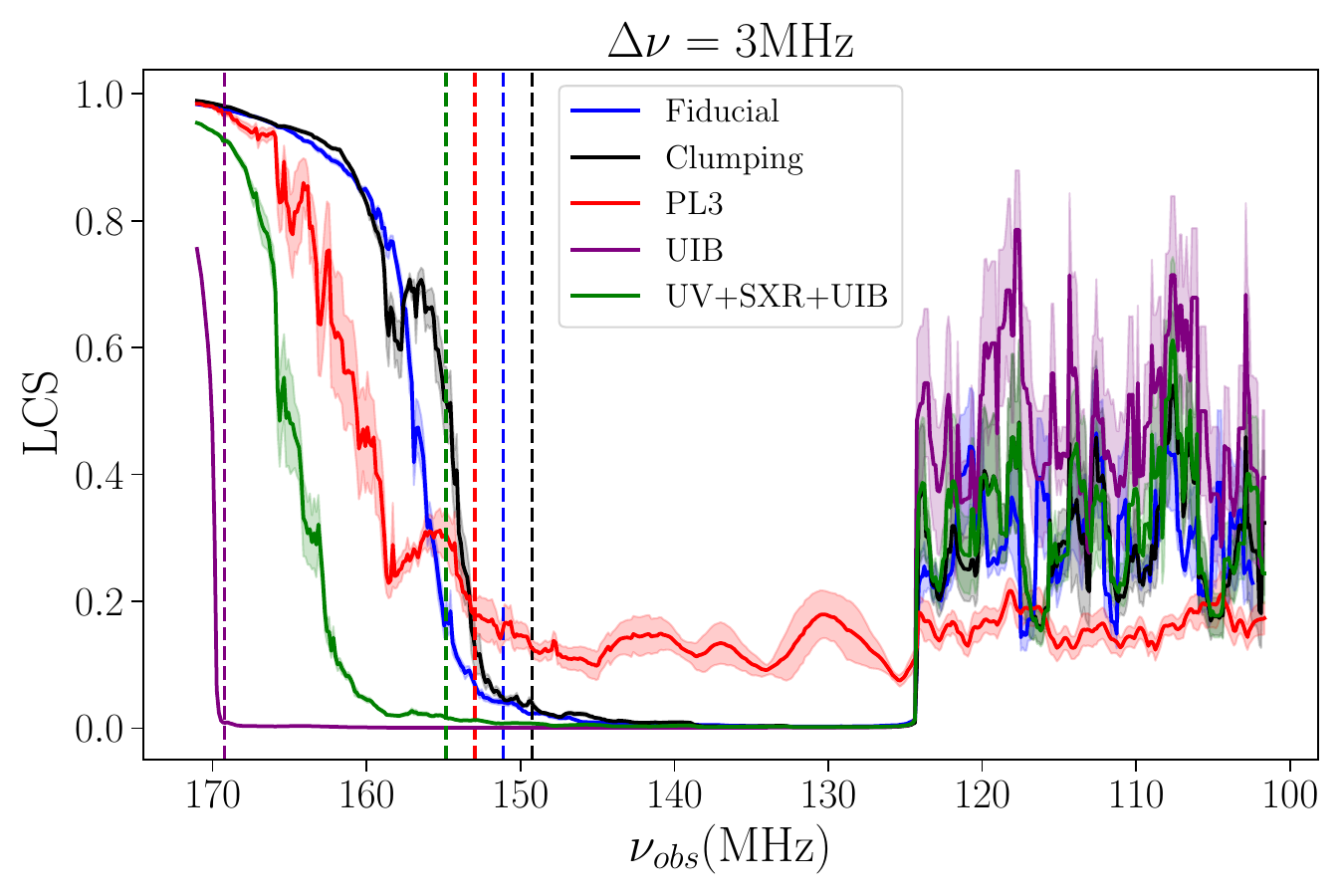}
    \end{minipage}
    \hfill
    \begin{minipage}[t]{0.49\textwidth}
        \centering
        \includegraphics[width=\linewidth]{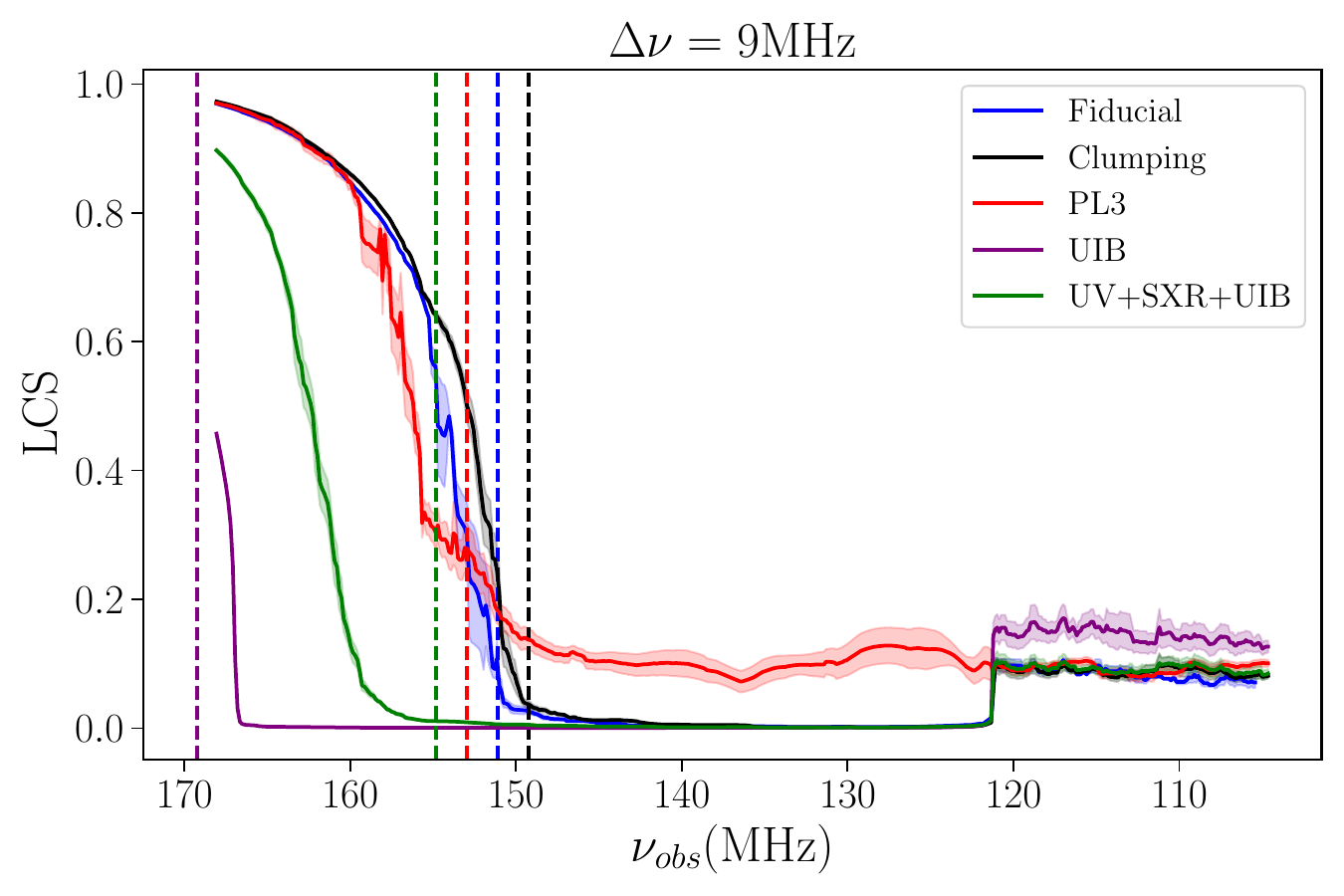}
    \end{minipage}
    \hfill
    \begin{minipage}[t]{0.49\textwidth}
        \centering
        \includegraphics[width=\linewidth]{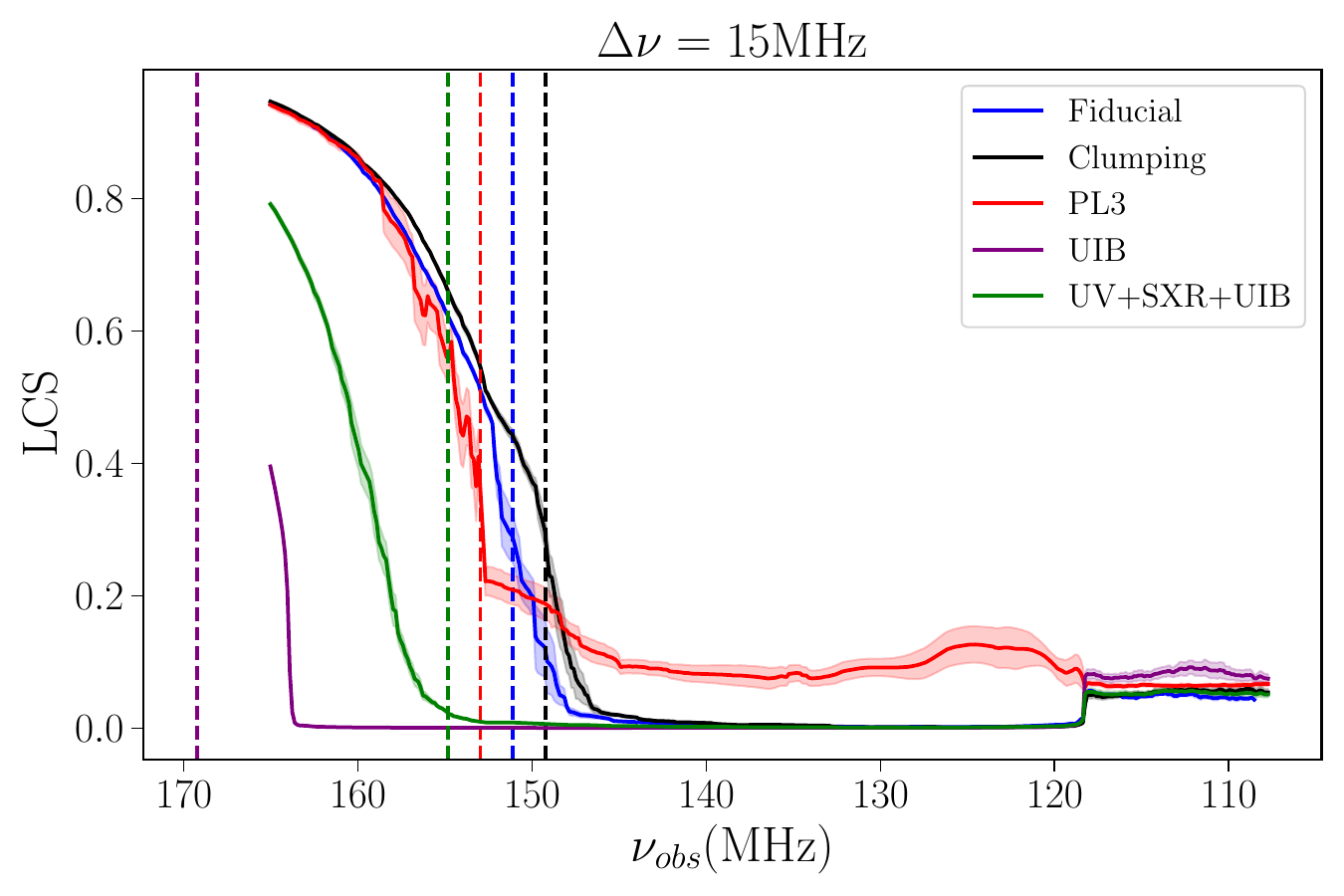}
    \end{minipage}
    \caption{Evolution of the LCS of moving volumes with their observed central frequency $\nu_{\rm obs}$ along lightcones of different reionization scenarios at spectral bandwidths $\Delta\nu =$ $3$ MHz, $9$ MHz, and $15$ MHz. The vertical dashed lines of respective colors denote the observed frequency at which the percolation transition occurs for the full coeval map of the corresponding reionization scenario.}
    \label{LCS_all}
\end{figure}

In this section, we discuss the evolution of the LIR using moving volume bandwidths of $3$ MHz, $9$ MHz, and $15$ MHz on lightcone maps of different reionization scenarios described in Table \ref{tab:reionization_scenarios}. The reionization scenarios considered in this work take into account different source properties and characteristics of the IGM, resulting in distinct topological evolution of HII bubbles. Figure \ref{LCS_all} shows the evolution of LCS with observed frequency in lightcones of the different reionization scenarios considered. The timing of the onset of percolation in the full coeval map of these reionization scenarios is indicated by the vertical dashed lines. As shown in Figure \ref{LCS_all}, each reionization scenario has a particular characteristic shape in the evolution of LCS with observed frequency, indicating that the percolation transition occurs at different observed frequencies. Additionally, the percolation transition shifts to lower observed frequencies with increasing moving volume bandwidth, similar to the fiducial model as shown in Figure \ref{fig:lcs_fidu}. The high fluctuations for smaller moving volume bandwidths are attributed to the lack of a sufficient number of ionized regions for a statistically significant estimation of LCS, and the discontinuity of interconnections in the subvolumes generated using the moving volume approach. As we increase the size of the subvolume, we account for an increasing number of interconnections, and it is reflected in the relatively smooth evolution of LCS.

The clumping model has similar source properties as those of the fiducial model, but different IGM characteristics, resulting in the formation of isolated neutral regions due to a non-uniform rate of recombination. These isolated HI regions lead to an increased number of tunnels in the ionization topology, with multiply connected ionized regions characterized by high genus values (see Figure 7, Paper I). At a fixed spectral bandwidth of moving volume, the onset of percolation in this model appears to happen at relatively lower observed frequencies compared to that of the fiducial scenario. Similar to the fiducial model, the onset of percolation in the clumping lightcone map at the moving volume bandwidth of 9 MHz matches that of the full coeval map, making it the optimal chunk size of this reionization scenario.

\begin{figure}[htbp]
    \centering
    \includegraphics[width=\linewidth]{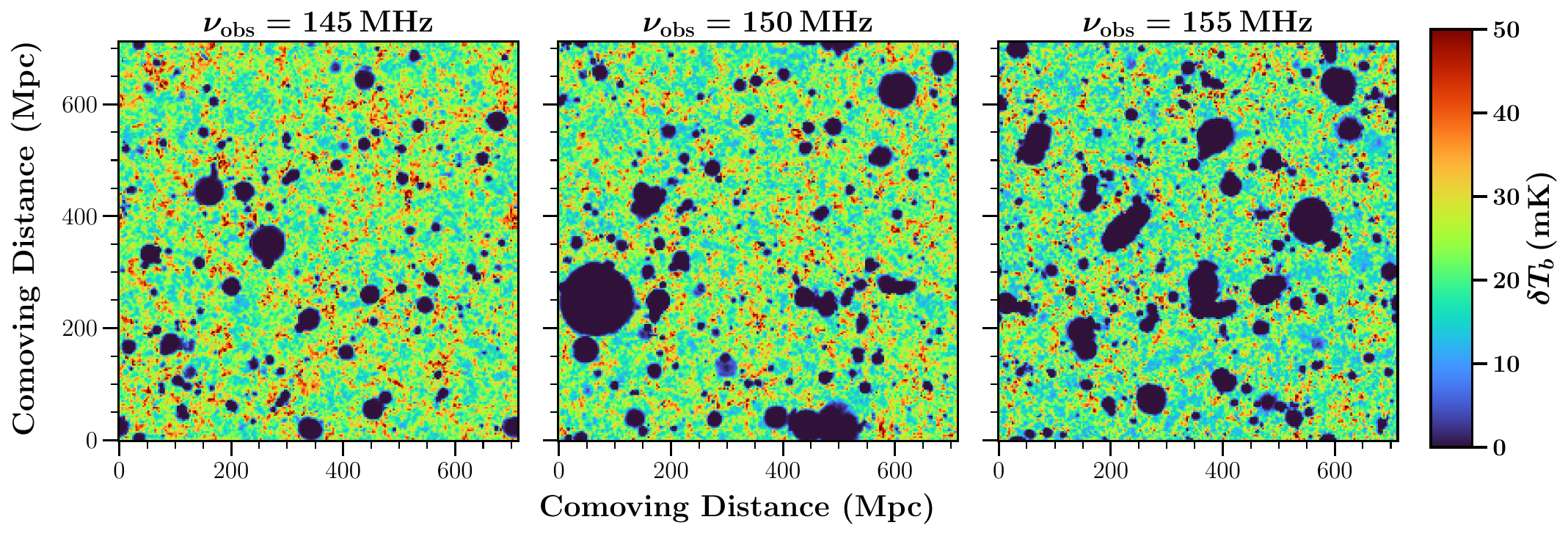}
    \caption{Slices of the PL($n=3$) model lightcone perpendicular to the line of sight at $\nu_{\rm obs} = 145$ MHz, $150$ MHz, and $155$ MHz.}
    \label{fig:pl3_slices}
\end{figure}

In the case of the PL($n=3$) model, the number of ionizing photons emitted by the sources residing in high-mass halos is much higher when compared to the other models, resulting in the formation of very large ionized bubbles around the source. As the lightcone map captures the spatial and temporal evolution of the 21-cm signal, each subvolume chosen to compute the LCS has a different mass distribution of halos, resulting in the formation of LIR at different locations as reionization progresses (see Figure \ref{fig:pl3_slices}). This results in a change in the LIR accounted for to compute LCS with observed frequency in subvolumes until the onset of percolation, reflected in the high variance in its evolution compared to other models. The model also exhibits bias in the observed onset of percolation at $3$ MHz and $9$ MHz moving volume bandwidths, and the onset of percolation appears to happen at later stages of reionization compared to that of the fiducial scenario for all the moving volume bandwidths considered. The timing of percolation for the $15$ MHz moving volume bandwidth closely matches the coeval case.

In the UIB model, reionization is dominated by hard X-rays, leading to outside-in reionization and a uniform background of ionization. This results in a highly diffuse network of interconnected ionized gas, and the percolation in the coeval cube of this model happens abruptly at a more advanced stage of reionization compared to other reionization scenarios considered. When evaluated along a lightcone, larger moving volumes span a wider evolutionary time along the line of sight, and in these bigger chunks, the higher-redshift end remains too sparsely ionized and is disconnected from the percolated LIR during the later stages of reionization. While the high-redshift end of a large chunk is less ionized across all reionization models, the UIB model is more affected due to the diffuse ionized region topology and abrupt percolation transition. As a result, the 3 MHz moving volume size considered in this model matches the percolation transition of the full coeval map, while the 9 MHz and 15 MHz moving volumes introduce a bias in the inferred transition.  

The UV+SXR+UIB model takes into account both inside-out and outside-in reionization processes. In this reionization scenario, like in other scenarios, the onset of percolation is determined by the bandwidth of the moving volume chosen for the analysis. The onset of percolation for the moving volume size of 15 MHz coincides with that of the coeval maps, making it the optimal chunk size to capture the percolation process. At a fixed bandwidth, the LCS rises at relatively higher observed frequencies than in the fiducial, clumping, and PL($n=3$) scenarios, but well before the UIB scenario, reflecting the intermediate topology created by a combination of inside-out and outside-in reionization processes.

As discussed above, the inferred percolation transition is highly sensitive to the spectral bandwidth of the moving volume, and the optimal bandwidth that best recovers the coeval percolation transition varies across reionization scenarios. Additionally, at any fixed moving-volume spectral bandwidth, the evolution of LCS remains highly sensitive to the underlying reionization scenarios with the same global ionization histories, making it a robust morphological statistic. Reionization scenarios can therefore be distinguished by the timing of the onset of percolation. This enables the theoretical models, which can be mapped directly onto lightcone predictions, to be compared against observations to determine the model governing cosmic reionization.
\subsection{Impact of instrumental effects and systematics}
\label{sec: noise_results}

We follow the procedure discussed in section \ref{sec:SKA_effects} to include the effect of beam and instrumental noise on the simulated maps. The added noise introduces random temperature fluctuations that shift the global minima of the mean-subtracted intensity maps, making it difficult to separate the ionized and neutral regions. In this analysis, we use Otsu's method \citep{1979Otsu} to identify the threshold that separates ionized regions from neutral ones. This is done by dividing a lightcone into subvolumes using the moving volume approach and using Otsu's method to choose a threshold to binarize these subvolumes. This method determines the optimal threshold by dividing the image into two classes and then maximizing the inter-class variance. Otsu's method is well-suited to the 21-cm signal as the brightness temperature distribution is expected to be bimodal during reionization, consisting of two classes, neutral and ionized regions.
It is important to note that the ability of Otsu's method to segment the image into two classes depends upon our assumption of $T_s \gg T_{\rm{CMB}}$. Spin temperature fluctuations are expected to lead to a much more continuous brightness temperature distribution, and Otsu's method is ineffective in regimes with the presence of substantial spin temperature fluctuations.

\begin{figure}[htbp]
    \centering
    \begin{minipage}[t]{0.49\textwidth}
        \centering
        \includegraphics[width=\linewidth]{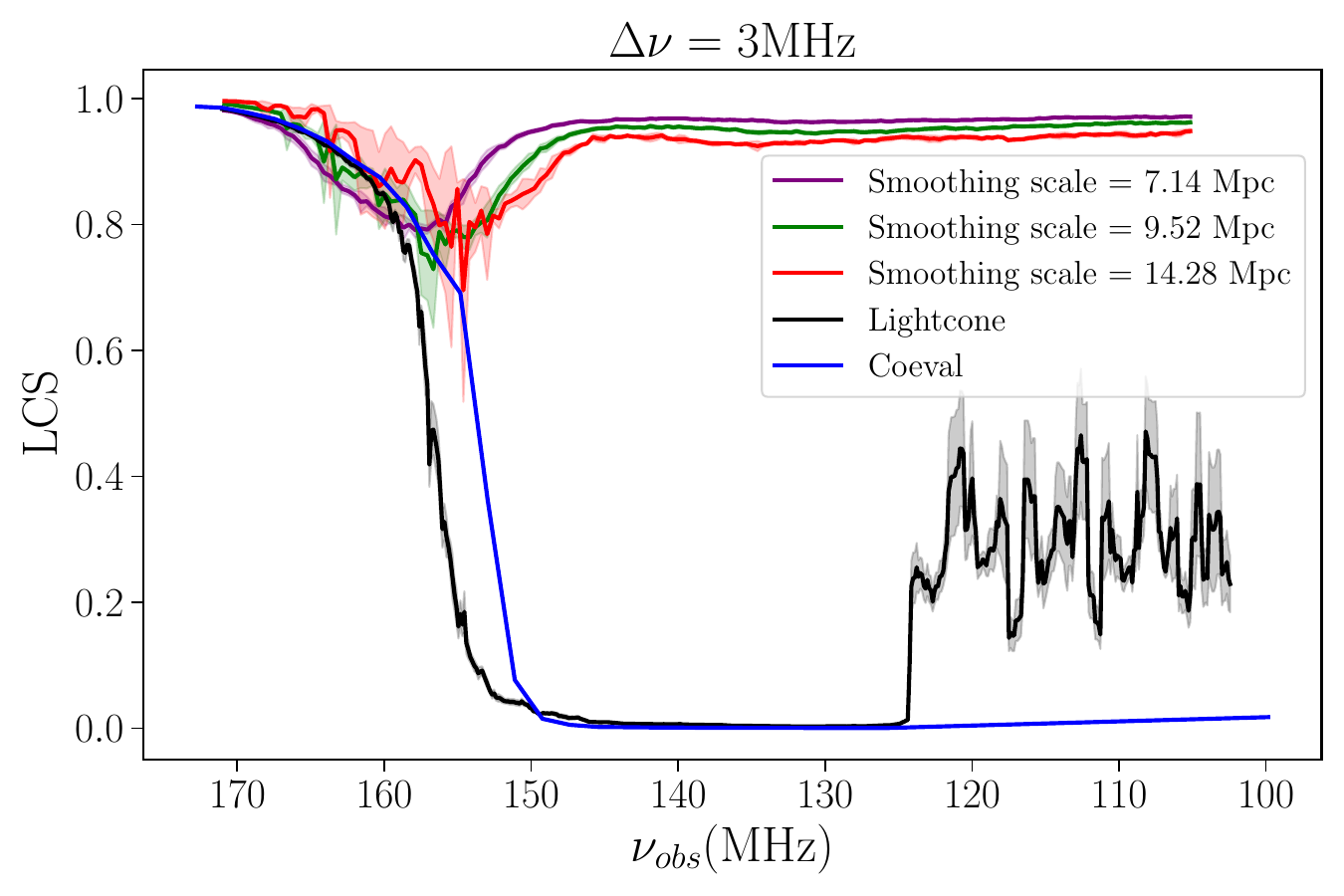}
    \end{minipage}
    \hfill
    \begin{minipage}[t]{0.49\textwidth}
        \centering
        \includegraphics[width=\linewidth]{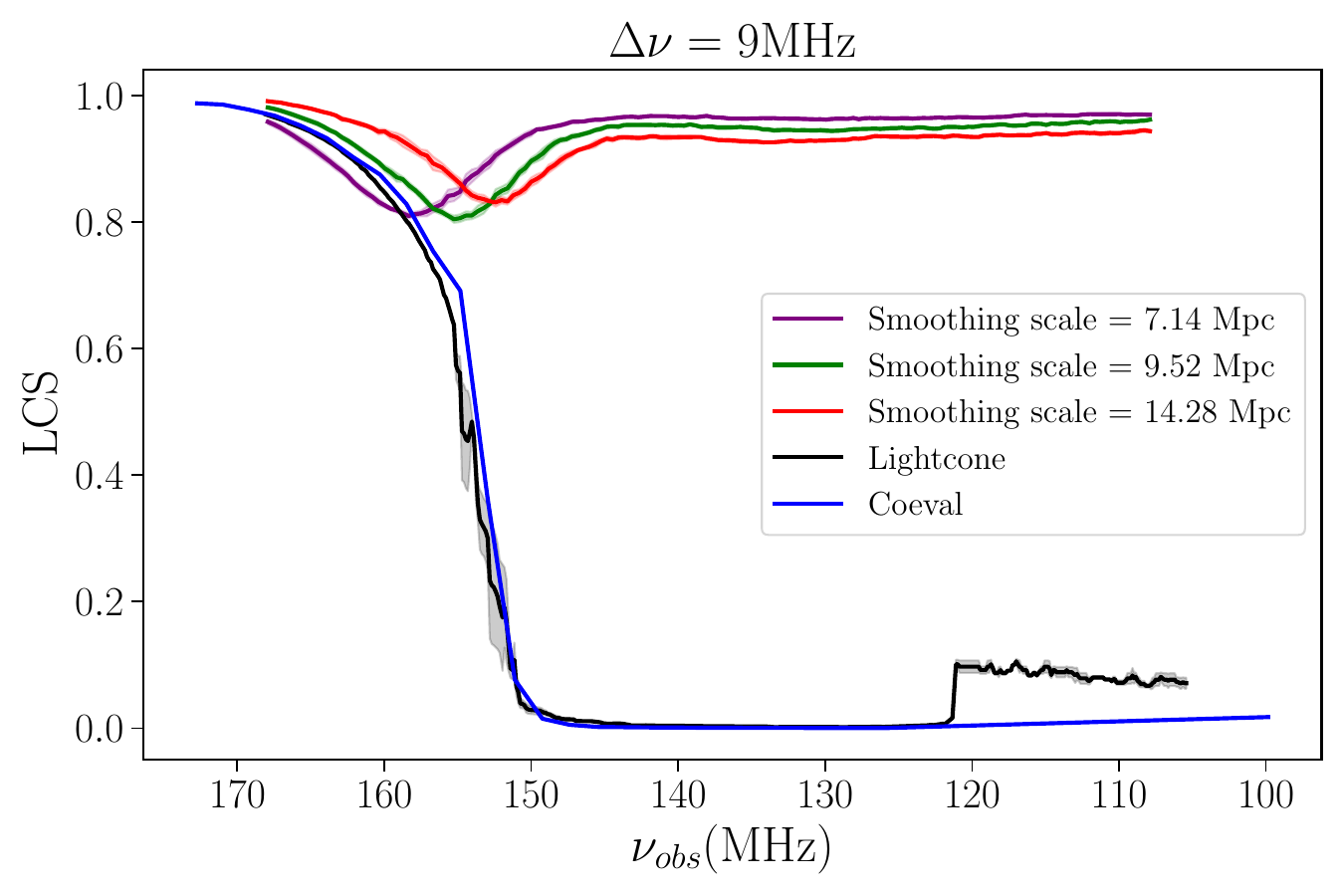}
    \end{minipage}
    \hfill
    \begin{minipage}[t]{0.49\textwidth}
        \centering
        \includegraphics[width=\linewidth]{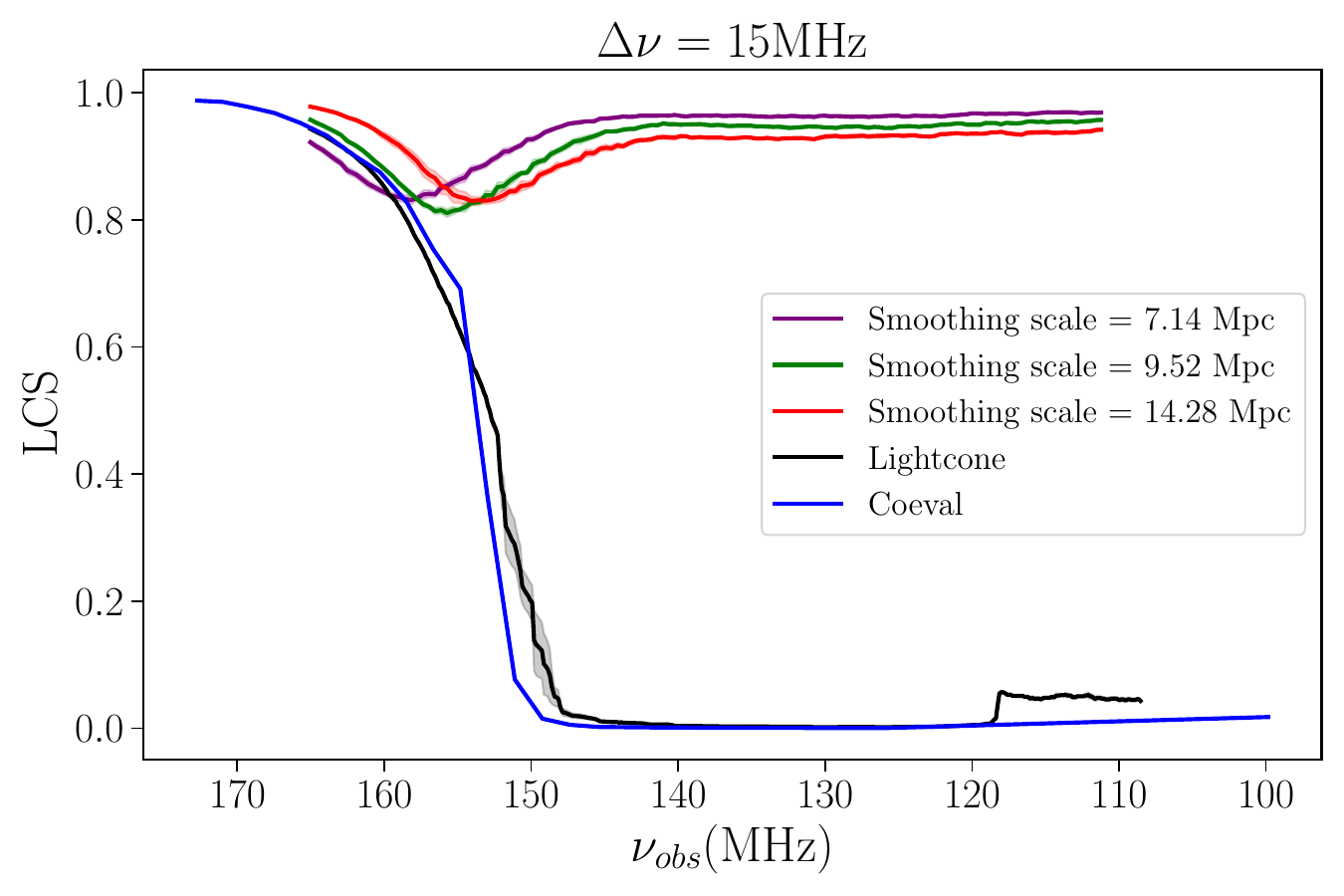}
    \end{minipage}
    \caption{Evolution of the Largest Cluster Statistic (LCS) as a function of the observed central frequency ($\nu_{\rm{obs}}$) for the fiducial lightcone evaluated using moving volume spectral bandwidths of $\Delta\nu = 3$, 9, and 15 MHz, and incorporating the effects of mock SKA-Low thermal noise alongside varying smoothing scales of 7.14, 9.52, and 14.28 Mpc. The blue curve represents the evolution in the full coeval volume of the fiducial reionization scenario and the black curve represents the evolution in the fiducial lightcone without any observational effects.}
    \label{LCS_noise_all}
\end{figure}

\begin{figure}[htbp]
    \centering
  
    \begin{minipage}[t]{0.49\textwidth}
        \centering
        \includegraphics[width=\linewidth]{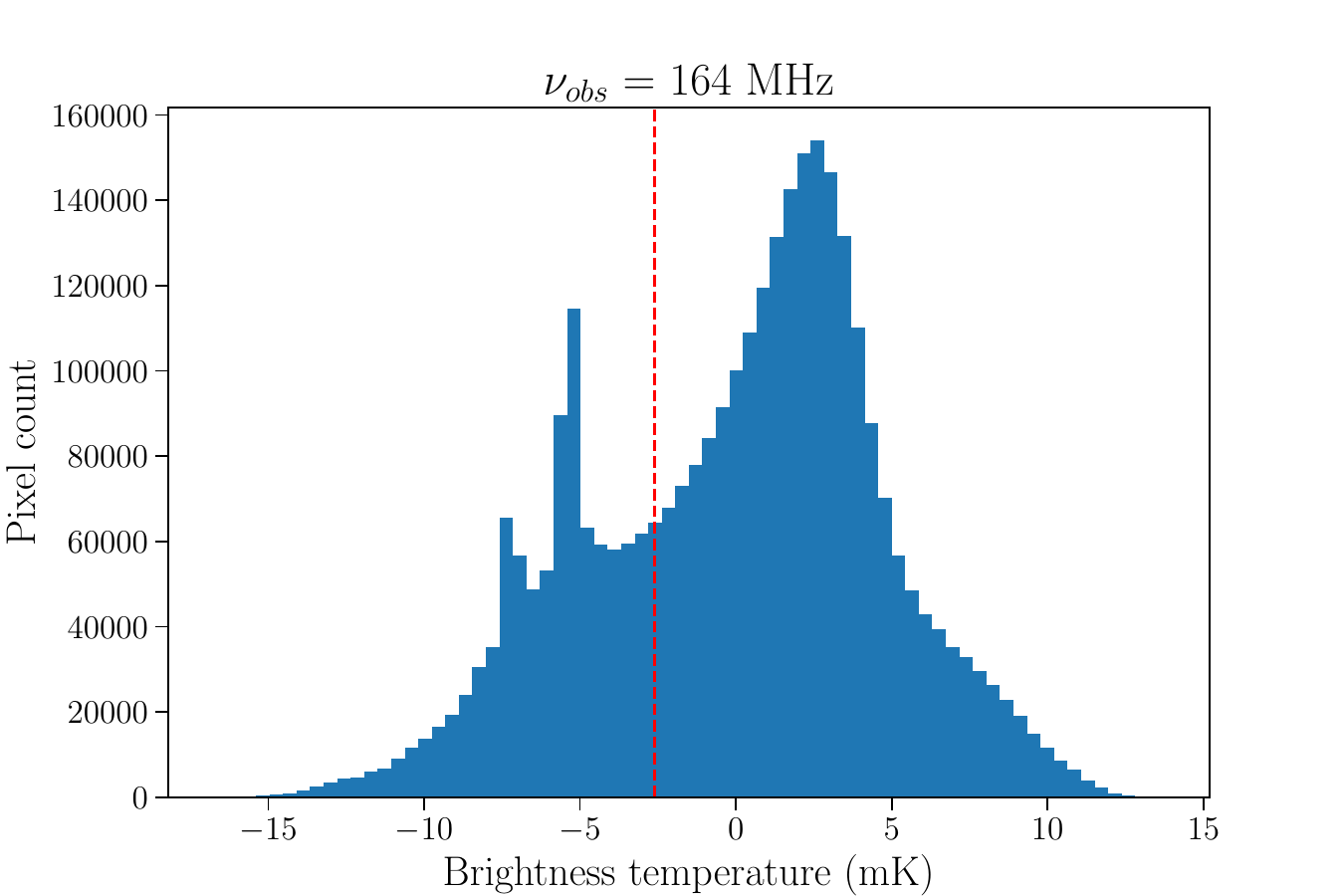}
    \end{minipage}
    \hfill
    \begin{minipage}[t]{0.49\textwidth}
        \centering
        \includegraphics[width=\linewidth]{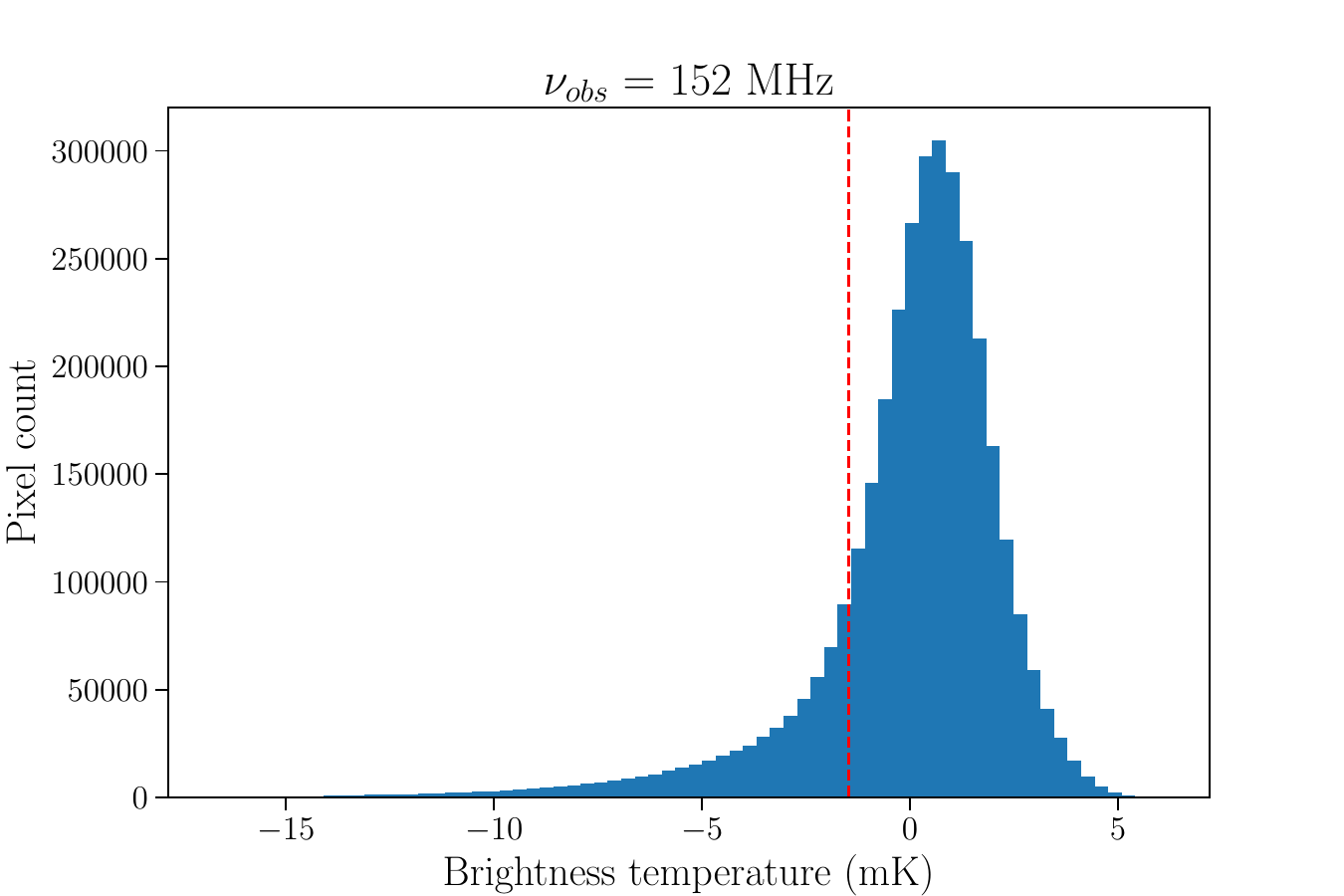}
    \end{minipage}
    \hfill
    \begin{minipage}[t]{0.49\textwidth}
        \centering
        \includegraphics[width=\linewidth]{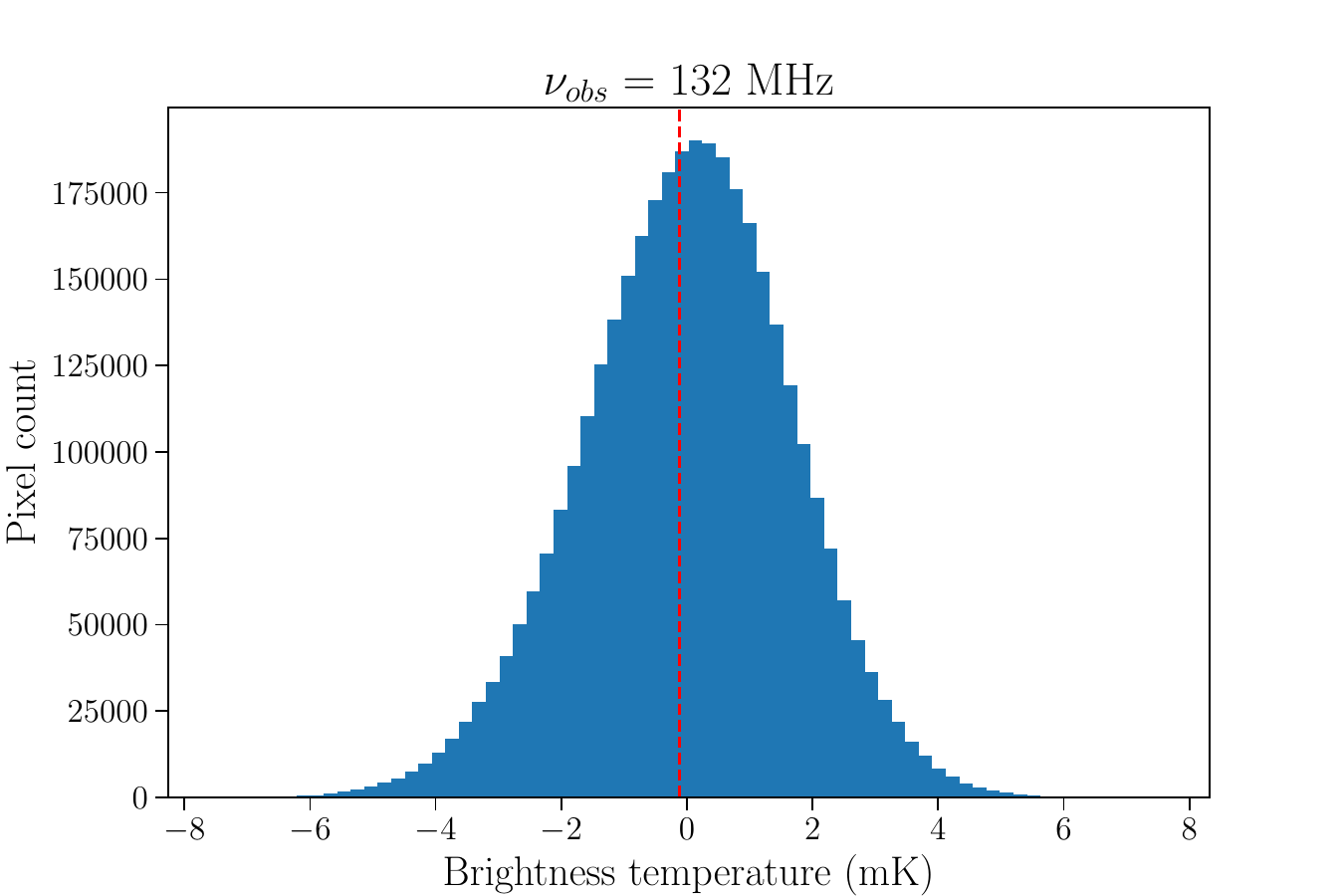}
    \end{minipage}
    \caption{Histograms of the 21-cm brightness temperature ($\delta T_b$) for the fiducial lightcone at observed frequencies of 164 MHz, 152 MHz, and 132 MHz, evaluated after applying SKA-Low thermal noise and a 2D Gaussian smoothing filter of size 9.52 Mpc. The vertical red dashed line in each panel indicates the optimal binarization threshold determined by Otsu's method. At the earlier stages of reionization (132 MHz), the distribution is noise-dominated and lacks a bimodal feature. At advanced stages of reionization (164 MHz), the ionized bubbles grow, and a distinct bimodal feature emerges.}
    \label{fig:otsu_hist}
\end{figure}

In Figure \ref{LCS_noise_all}, we plot the evolution of the LCS with observed frequency for different moving volume sizes and smoothing scales. At the earlier stages of reionization, the signal loses its bimodal feature, and instrumental thermal noise dominates the temperature fluctuations. At these frequencies, Otsu's method attempts to binarize the map by splitting the single peak into two classes and determines the threshold as shown in Figure \ref{fig:otsu_hist}. The pixels thus identified as ionized connect to form a cluster, artificially increasing the LCS. As reionization progresses, the ionized bubbles grow and a second peak begins to emerge. During this time, Otsu’s threshold moves out of the noise peak and into the developing valley between the two modes, resulting in a drop in LCS value. During the advanced stages of reionization, the signal exhibits a distinct bimodal feature allowing Otsu's method to determine a robust threshold between the ionized and neutral regions as shown in Figure \ref{fig:otsu_hist}. The transition from the noise-dominated regime to the signal-dominated bimodal regime is observed as a dip in the evolution of LCS for all moving volume sizes and smoothing scales.

Consistent with previous studies on coeval maps (e.g., Paper IIa, Paper IIb), we observe that the instrumental beam has a profound impact on the inferred percolation transition. As the smoothing length increases from 7.14 Mpc to 14.28 Mpc, the sharp rise in the LCS is shifted towards lower observed frequencies (earlier stages of reionization). This arises because Gaussian smoothing averages the 21-cm signal over the beam scale, creating patches of intermediate brightness temperature that mimic numerous partially ionized regions. At early times, when the ionized bubbles are small, the binarization scheme identifies most of them as neutral regions. At later times, as the overall ionized volume increases, the same averaging effect shifts these intermediate regions across the threshold determined by Otsu's method. With the increase in smoothing scale, the averaging effect creates more partially ionized regions at the intermediate stages of reionization, resulting in the shift of the transition dip in the percolation curves towards lower observed frequencies. Additionally, Figure \ref{LCS_noise_all} reveals that the choice of moving volume size introduces a relatively minimal effect in the evolution of LCS compared to the dominant bias introduced by the telescopic beam. This indicates that the FWHM of the instrumental beam dictates the recoverability of the percolation transition far more than the spectral bandwidth of the binarized map.

\begin{figure}[htbp]
    \centering
    \begin{minipage}[t]{0.49\textwidth}
        \centering
        \includegraphics[width=\linewidth]{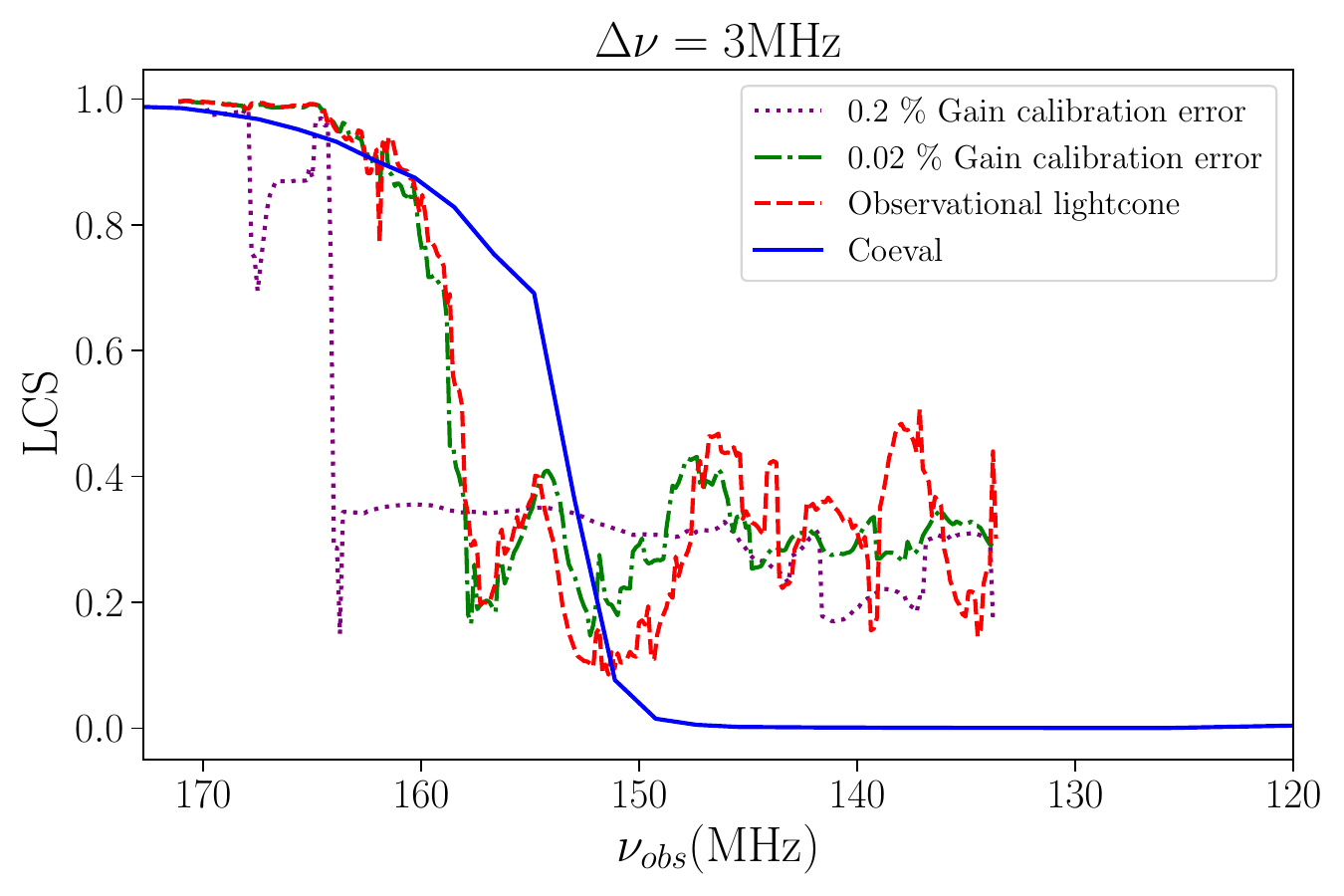}
    \end{minipage}
    \hfill
    \begin{minipage}[t]{0.49\textwidth}
        \centering
        \includegraphics[width=\linewidth]{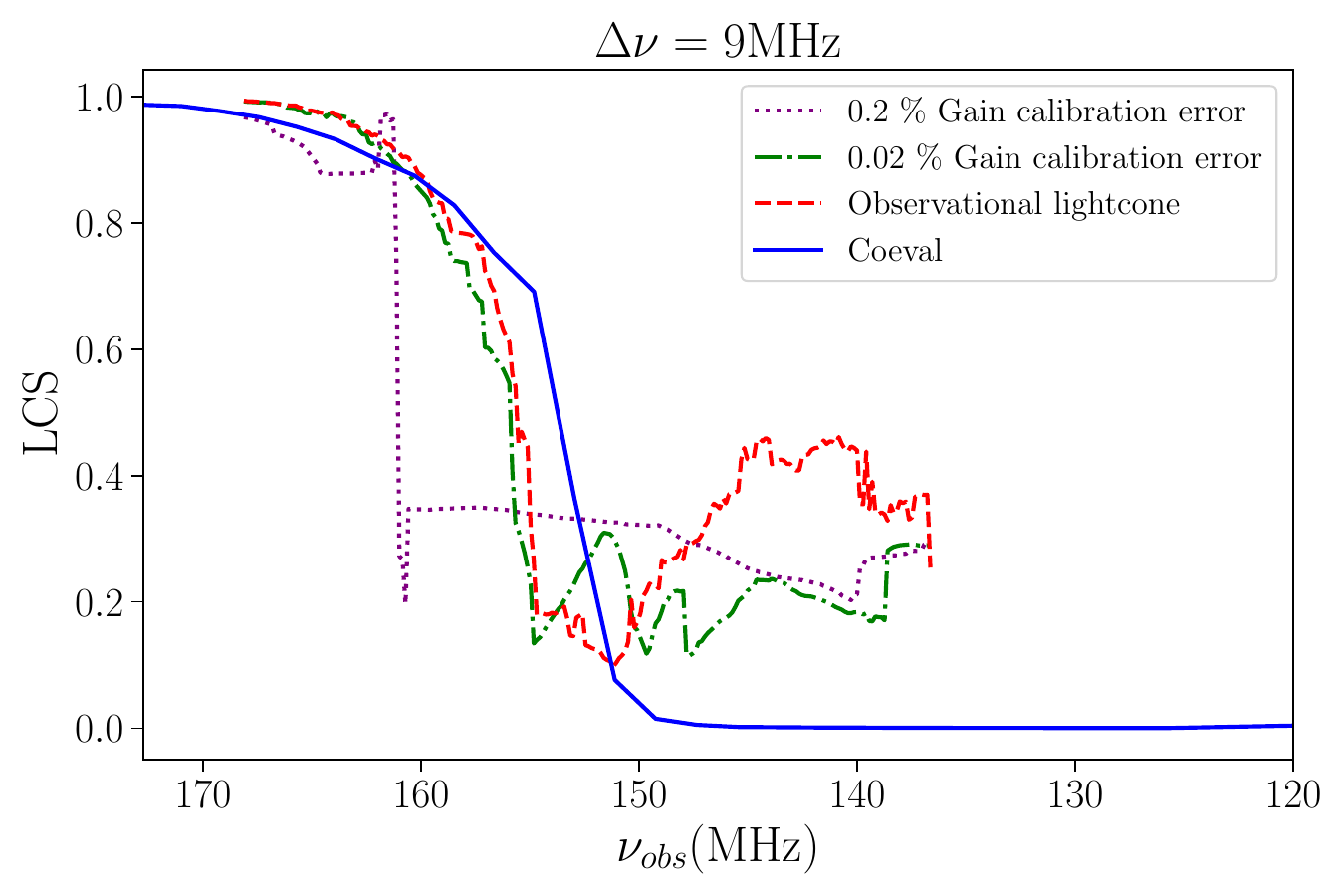}
    \end{minipage}
    \hfill
    \begin{minipage}[t]{0.49\textwidth}
        \centering
        \includegraphics[width=\linewidth]{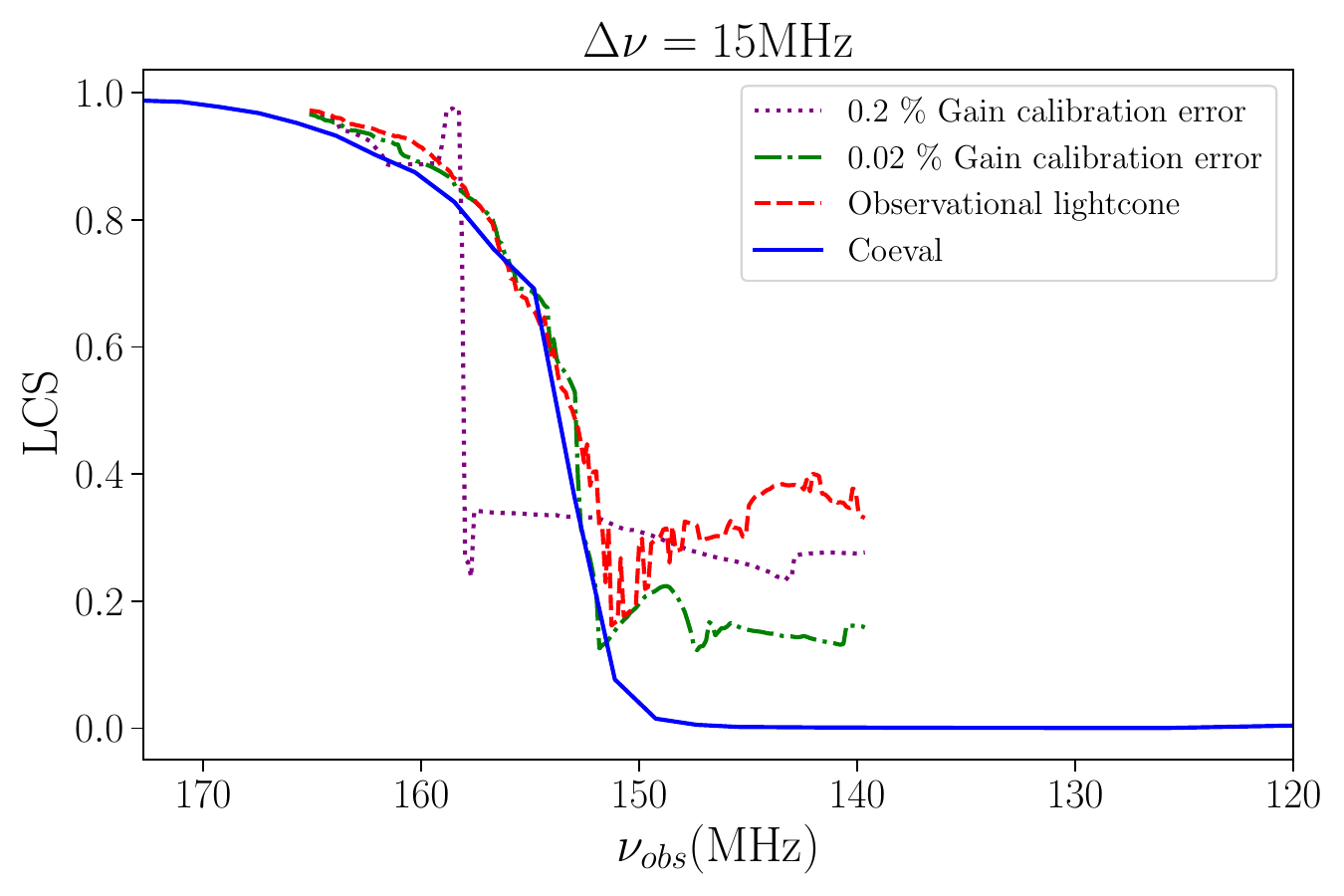}
    \end{minipage}
    \caption{Evolution of the obtained LCS from simulated SKA-Low 21\,cm observation as a function of the observed central frequency ($\nu_{\rm{obs}}$) for the fiducial lightcone evaluated using moving volume spectral bandwidths of $\Delta\nu = $ 3, 9, and 15 MHz, and incorporating the effects of antenna-based gain calibration errors along with 21\,cm signal without any contamination. The blue curve represents the evolution in the full coeval volume of the fiducial reionization scenario.}
    \label{LCS_calib_all}
\end{figure}

In Figure \ref{LCS_calib_all}, we plot the evolution of the LCS obtained from the 21\,cm lightcone with observed frequency for different gain calibration errors, along with that from the 21\,cm lightcone without any contamination from residual systematic effects. To binarise the observed differential brightness temperature map, we use adaptive thresholding based on the Yen method \cite{YenMethod}. Instrumental systematics arising from residual calibration errors, along with non-linear deconvolution artifacts, randomly shift pixel values. This makes a simple global thresholding method (such as Otsu's method) inefficient at distinguishing neutral and ionized regions. Since the target signal brightness-temperature distribution evolves with redshift (along the line of sight), we apply two channel-dependent thresholding criteria. In radio interferometric observations, we measure only the fluctuations in the brightness temperature field. Accordingly, the minima of the brightness temperature maps correspond to ionized regions, as expected. Therefore, we enforce the physical constraint that pixels with positive brightness temperature are never classified as ionized. While a negative value is a necessary condition for ionization, it is not sufficient, as thermal noise and low-density fluctuations in neutral regions can also yield negative brightness temperatures. The thresholding scheme described below determines which negative pixels correspond to true ionized regions. We first smooth 21-cm maps with a 2D Gaussian kernel to suppress pixel-scale noise. Then it is thresholded using Yen's method, applied differently depending on whether neutral or ionized regions dominate the frequency channel. For the neutral-dominated channels, the threshold is computed from negative pixels only to avoid classifying faint neutral gas as ionized. In ionized-dominated channels, it is computed from the histogram and capped at zero, so the sign constraint still holds.

As discussed in paper IIb, the gain calibration error level of $0.2 \%$ introduces severe spatial and spectral fluctuations that obscure the 21-cm signal. This artifact contamination results in an artificial fragmentation of the LIR, resulting in the erratic evolution of the LCS shown in Figure \ref{LCS_calib_all}. However, an optimistic gain calibration error level of $0.02 \%$ recovers the shape of the percolation transition across all moving volume bandwidths, closely following the observed 21-cm signal without any contamination (observational lightcone). At this optimistic tolerance level, the spatial artifacts introduced by calibration inaccuracies remain sub-dominant (see paper IIb), and a 15 MHz moving volume serves as the optimal chunk size to determine the onset of percolation for both the observational lightcone and the optimistically calibrated maps. It is important to note that this evolution is sensitive to the thresholding method chosen to binarize the maps.
\section{Summary}
\label{sec:summary}

The progression of reionization of HI gas in the intergalactic medium and the sources of ionizing photons that led to it are not well understood to date. Possible sources of reionization include UV radiation from young stars within galaxies, hard X-rays emitted by AGNs that form a uniform ionizing background, soft X-ray photons originating from high-redshift starburst galaxies, X-ray binaries, etc. The relation between the rate of emission of ionizing photons by a source and its host halo mass is also not well understood, and there can be multiple possible relations between the two quantities. Additionally, the rate of recombination of HII is expected to be density dependent. All the above factors lead to multiple different reionization scenarios, of which we studied five scenarios as detailed in \ref{sec:reionization_scenarios}.

In order to interpret tomographic 21-cm intensity maps from the EoR and test the plausibility of the different reionization scenarios, the signal needs to be forward modeled and compared with observations. For this, summary statistics that both reduce the dimensionality of the data and, at the same time, capture the key information of the signal are needed. Image-based statistics, such as LCS, are helpful for this purpose, as they capture the topological information of ionized regions. The ability of LCS to distinguish between these scenarios based on the timing of their percolation transition (the stage at which LCS rises sharply towards unity) was studied in detail in Paper I. In this work, we have extended this analysis to more realistic observations by incorporating the lightcone effect into our simulations, which is inherent to intensity mapping experiments over a wide frequency (redshift) range.

Estimating the LCS of a 21-cm lightcone would yield a single value, washing out the information of its redshift evolution, and hence the percolation transition. In order to capture the redshift evolution of LCS, a lightcone must be divided into subvolumes, and the LCS must be computed on these subvolumes to characterize its evolution. The subvolume must be small enough such that the LCS is not averaged over a wide redshift range. At the same time, it must be large enough so that it accommodates a large number of ionized regions, and the computed LCS value is statistically significant. Dividing the lightcone into such disjoint subvolumes would give a small number of them, and the redshift evolution of LCS would not be sampled effectively. We therefore employed a moving volume approach where the subvolume translates along the lightcone as described in \ref{sec:LCS_on_lightcones}.

The inferred percolation transition is highly sensitive to the spectral bandwidth of the moving volume considered to compute the LCS. As we increase the size of the moving volume, the timing of the onset of percolation appears to happen at the earlier stages of reionization. The optimal moving volume bandwidth required to accurately determine the onset of percolation depends on the reionization scenario considered. In the case of the fiducial and clumping scenarios, a moving volume bandwidth of 9 MHz matches the percolation transition estimated from coeval boxes, whereas in the PL3 scenario, where the larger ionized bubbles are generated, a moving volume bandwidth of 15 MHz is considered to be optimal to understand the percolation transition. Additionally, the optimal bandwidth also varies significantly when outside-in source models are considered. The highly diffuse topology of the UIB scenario requires a narrow 3 MHz spectral bandwidth to prevent discontinuity of interconnections along the line of sight, whereas a composite UV+SXR+UIB model is optimally evaluated with a moving volume spectral bandwidth of 15 MHz. Additionally, at any fixed bandwidth, the timing of the onset of percolation varies among the scenarios considered. In the case of inside-out reionization scenarios like fiducial, clumping, and PL($n=3$), HII regions percolate at lower observed frequencies. Including the outside-in reionization processes delays the onset of percolation as observed in UIB and UV+SXR+UIB models. This differential evolution enables LCS computed at fixed spectral bandwidth to be used as a summary statistic in the forward modeling inference framework to distinguish between reionization scenarios.

When extending our analysis to incorporate realistic SKA-Low instrumental effects, we find that the percolation transition is predominantly dictated by the FWHM of the instrumental beam rather than the choice of spectral bandwidth. The spatial resolution, governed by the instrumental beam's full width at half maximum (FWHM), spatially averages the 21-cm signal and creates numerous partially ionized regions. Global binarization algorithms used to identify the ionized regions, like Otsu's method, misclassify these partially ionized regions during the earlier stages of reionization. Despite this, the evolution of LCS exhibits a characteristic dip prior to percolation, marking the transition from a noise-dominated regime to a signal-dominated bimodal regime, and a subsequent rise driven by the percolation of HII bubbles. This distinct feature in the LCS evolution enables the determination of the timing of the percolation in the planned SKA-Low observations. Furthermore, the introduction of residual antenna-based gain calibration errors randomly shifts pixel values, rendering global thresholding highly inefficient. By implementing an adaptive, channel-dependent binarization approach based on Yen's algorithm, a $0.2 \%$ error introduces severe artifacts that artificially fragment the LIR and yield an erratic LCS evolution. While an optimistic gain calibration error of $0.02 \% $, allows a 15 MHz moving volume to successfully recover the timing of the onset of percolation. 

Given the extreme sensitivity of topological recovery to these binarization algorithms, future image-based analyses of the EoR will critically require the development of more robust thresholding algorithms capable of accurately isolating ionized regions in heavily contaminated observational datasets. Furthermore, we plan to extend this framework to more realistic and physically motivated reionization scenarios, and incorporate additional observational systematics, such as foreground residuals and ionospheric effects, to improve our understanding of their implications for HII percolation in EoR 21-cm observations.



\acknowledgments

The authors would like to acknowledge useful discussions with Ilian T. Iliev regarding this work. The authors also thank the anonymous referee for their constructive comments and insightful observations, which helped improve this manuscript. HP would like to acknowledge the funding support provided by Stellenbosch University and the University of Groningen. MMD, SM, and AD thank the Science and Engineering Research Board (SERB) and the Department of Science and Technology (DST), Government of India, for financial support through Core Research Grant No. CRG/2021/004025 titled “Observing the Cosmic Dawn in Multicolor using Next Generation Telescopes”. CSM would like to acknowledge financial support from the Council of Scientific and Industrial Research (CSIR) via a CSIR-SRF Fellowship (Grant No. 09/1022(0080)/2019-EMR-I) and from the ARCO Prize Fellowship. SKP acknowledges the financial support from the Department of Science and Technology, Government of India, through the INSPIRE Fellowship [IF200312]. LN acknowledges the financial support from the Department of Science and Technology, Government of India, through the INSPIRE Fellowship [IF210392]. The authors acknowledge the use of facilities procured through funding via the Department of Science and Technology, Government of India-sponsored DST-FIST grant no. SR/FST/PSII/2021/162(C) awarded to the DAASE, IIT Indore.


\bibliographystyle{JHEP}
\bibliography{references}


\end{document}